%% file: main.tex
\pdfoutput=1
\documentclass[10pt,twocolumn,letterpaper]{article}

\usepackage[pagenumbers]{cvpr} % To force page numbers, e.g. for an arXiv version

\input{preamble}
\definecolor{cvprblue}{rgb}{0.21,0.49,0.74}
\usepackage[pagebackref,breaklinks,colorlinks,allcolors=cvprblue]{hyperref}

\def\paperID{13888} % *** Enter the Paper ID here
\def\confName{CVPR}
\def\confYear{2026}

\title{Improving the Plausibility of Pressure Distributions Synthesized from Depth Image through Generative Modeling}

\author{
Neevkumar Manavar$^{1}$ \quad
Hanno Gerd Meyer$^{1}$ \quad
Joachim Waßmuth$^{1}$ \\
Barbara Hammer$^{2}$ \quad
Axel Schneider$^{1}$
\vspace{5pt} \\
$^{1}$Hochschule Bielefeld, Interaktion 1, Bielefeld, Germany \\
$^{2}$CITEC, Bielefeld University, Inspiration 1, Bielefeld, Germany \\
\vspace{5pt}
{\tt\small \{neevkumar.manavar, hanno\_gerd.meyer, joachim.waßmuth, axel.schneider\}@hsbi.de} \\
{\tt\small bhammer@techfak.uni-bielefeld.de}
}

\usepackage{multirow}
\usepackage[nolist]{acronym}

\usepackage{float}
\floatstyle{plaintop}
\restylefloat{table}

\usepackage{tabularx}

\usepackage{amsmath}

\usepackage[export]{adjustbox}

\usepackage{array}  

\newcolumntype{Y}{>{\centering\arraybackslash}X}

\usepackage{siunitx}

\begin{document}
\maketitle
\input{sec/0_abstract}    
\input{sec/1_intro}
\input{sec/2_related_work}
\input{sec/3_methods}

\input{sec/4_evaluation}

\input{sec/5_ablation_study}
\input{sec/6_conclusion}
{
    \small
    \bibliographystyle{ieeenat_fullname}
    \bibliography{main}
}

\input{sec/X_suppl}

\input{sec/7_abbreviations}

\end{document}

%% file: sec/0_abstract.tex
\begin{abstract}

Monitoring contact pressure in hospital beds is essential for preventing pressure ulcers and enabling real-time patient assessment. Current methods can predict pressure maps but often lack physical plausibility, limiting clinical reliability. This work proposes a framework that enhances plausibility via \acf{ILS} and \acf{WOL} with conditional generative modeling to produce high-fidelity, physically consistent pressure estimates. This study also applies diffusion based conditional \acf{BBDM} and proposes training strategy for its latent counterpart \acf{LBBDM} tailored for pressure synthesis in lying postures. Experiment results shows proposed method improves physical plausibility and performance over baselines: \acs{BBDM} with \ac{ILS} delivers highly detailed maps at higher computational cost and large inference time, whereas \acs{LBBDM} provides faster inference with competitive performance. Overall, the approach supports non-invasive, vision-based, real-time patient monitoring in clinical environments.

\end{abstract}

%% file: sec/1_intro.tex
\section{Introduction}
\label{sec:intro}

\begin{figure*}[t]
  \centering
  \includegraphics[width=0.99\linewidth]{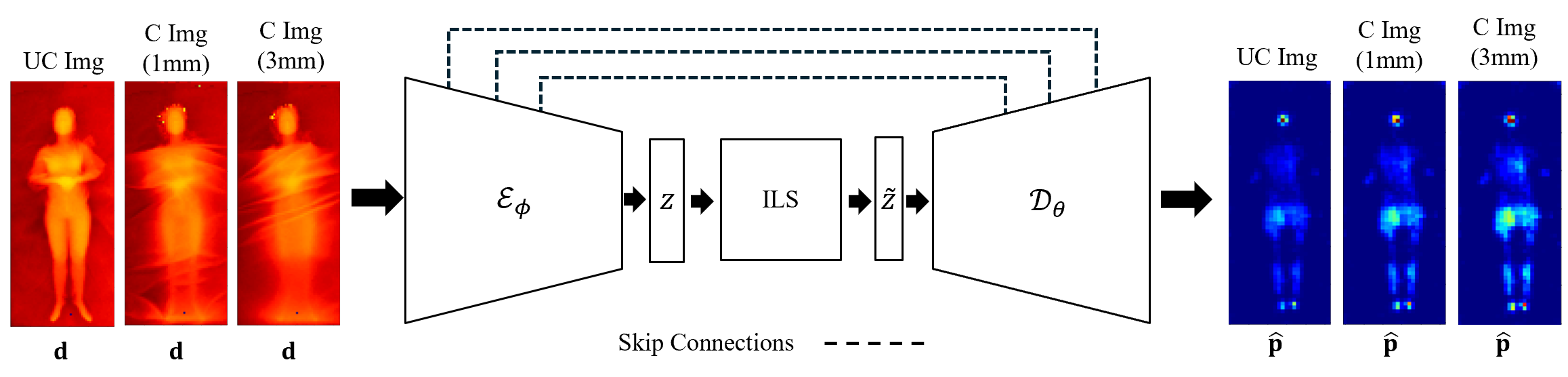}
  \caption{Overview of the model architecture with the integration of the ILS module. The model takes a single depth image as input and generates the pressure distribution.}
  \label{fig:ILS_overview}
\end{figure*}

The global prevalence of hospitalized adult patients with pressure injuries is 12.8\%, with an incidence rate of 5.4 per 10,000 patient days \cite{LI2020103546}. Mechanical factors such as shear force, friction, and pressure remain the leading risk contributors \cite{EPUAP2019}. Several wearable medical devices have been proposed to monitor patients' risk of pressure ulcer development \cite{WearableDevice, lee_fabric-based_2023}; however, these devices often contribute to pressure-related injuries \cite{MedicalDevicePU, JACKSON_Pressure_Injury_Surevey, Kayser2018_MDRPI}.

To prevent pressure injuries caused by mechanical boundary conditions, repositioning patients every two hours is a commonly used practice \cite{Defloor2006}. However, continuous contact pressure monitoring of lying patients is critical for understanding, prevention, and timely intervention against pressure injuries \cite{Osuagwu2023PM, GHOSH2024100029}. Beyond traditional active pressure monitoring systems \cite{MonitorPressurePoints, PressureMonitoringSleep}, recent studies have explored predicting contact pressure distributions using non-invasive, vision-based approaches. For instance, \citeauthor{CleverBodyPressure} \cite{CleverBodyPressure} and \citeauthor{ManavarAttFnet} \cite{ManavarAttFnet} predict 2D pressure distributions of individuals lying in bed using depth images.

The method by \citeauthor{CleverBodyPressure} \cite{CleverBodyPressure} employs a deep neural network encoder followed by a white-box reconstruction model for pressure estimation. In contrast, \citeauthor{ManavarAttFnet} \cite{ManavarAttFnet} introduced the \ac{AttnFnet} with a \ac{cGAN} framework to synthesize pressure distributions. While these approaches outperform baselines, problem of physically implausible distributions persist \cite{ManavarAttFnet}.

To address this issue, we propose an improved framework for generating plausible pressure distributions. The study introduces the \acf{ILS} and \acf{WOL} modules, to tackle this problem. We build upon the AttnFNet architecture as the base model, which encodes depth images into a latent space, while anthropometric factors such as mass, height, and gender are conditioned through cross-attention within this space. During training, the encoded mass information is further optimized using \ac{WOL}. 

A major limitation of prior methods is their weak modeling of how pressure depends on human attributes (mass, height, gender). The proposed \ac{ILS} addresses this by explicitly coupling these factors, producing more plausible and adaptive pressure maps. A practical advantage is the model's body parameter adaptation: for the same subject's depth image, adjusting mass, height, or gender yields distinct, physically consistent pressure distributions. This adaptability is clinically useful when patient characteristics change over time, since only the conditioning inputs need updating rather than retraining the model.

This study also introduces pressure synthesis via a conditional \acf{BBDM} and \acf{LBBDM} \cite{Li_2023_BBDM}. We employed a two-stage training strategy for a conditional \ac{LBBDM}, tailored for pressure synthesis. Additionally, a VQGAN model \cite{VQGAN2022} trained on a distinct objective is integrated with the \ac{LBBDM} for pressure generation. To see the impact of \ac{ILS} in the diffusion process, we have incorporated \ac{ILS} into the denoising model and into the pretrained \ac{AttnFnet} network when using conditional \ac{LBBDM}.

We evaluate the proposed method on the publicly available \ac{SLP} dataset \cite{LiuSLP} and compare it against established baselines \cite{CleverBodyPressure, ManavarAttFnet}.

% Recent development in patient fall detection, sensored human bed to regularised back pressure became backbone for patient monitoring. In bed human pose estimation and pressure prediction further enhance the visualization of patience motion. 

%% file: sec/2_related_work.tex
\section{Related Work}
\label{sec:related_work}

\textbf{Pressure Injury Prevention:} According to the European Pressure Ulcer Advisory Panel (EPUAP) \cite{EPUAP2019}, mechanical boundary conditions are one of the causes of pressure ulcer development. Injury formation occurs due to boundary conditions such as the magnitude, duration, and time of the applied mechanical load. Patients with mobility impairments are particularly vulnerable to such conditions.

Conventionally, nurses reposition bedridden patients every two hours to reduce skin hydration, minimize prolonged mechanical load, and redistribute pressure \cite{PressureInjuryPreventionSurvey}. As this repositioning frequency dates back to World War II \cite{Defloor2006}, it remains a standard method for preventing pressure injuries caused by mechanical boundary conditions. However, studies by \citeauthor{PatientRepositioning} \cite{PatientRepositioning, EffectsOfTurning} suggest that the current repositioning protocol requires improvement, as poor pressure redistribution can still occur despite the scheduled turning.

In light of this, inferring accurate pressure distributions for a lying person is essential to understanding ulcer formation caused by mechanical factors and enabling timely intervention \cite{Osuagwu2023PM, GHOSH2024100029}. Building upon this concept, this paper addresses the problem through a non-invasive, lightweight, vision-based system that infers biomechanical contact pressure distributions from a single depth camera.

\textbf{Patient Monitoring Systems and Pressure Synthesis:} The most common approach for monitoring biomechanical contact pressure is through direct sensing devices \cite{MonitorPressurePoints, PressureMonitoringSleep}, which have been improved by techniques such as body part localization \cite{BodypartLocalization} and automatic limb identification \cite{AutomaticLimbIdentification} to aid pressure injury prevention.

Several studies have focused on classifying the postures of lying humans \cite{PostureLyingHuman, PostureFuzzyNetwork, PostureMultiViewpoints}. Later, \citeauthor{LiuSLP} \cite{LiuSLP} introduced the \ac{SLP} dataset, enabling in-bed human pose monitoring \cite{PyramidFusion, Muller_Pose_CVPR, SLPGANPose, PrivacyPose, Accurate_Pose, CleverPose3D}.

Alternatively, some researchers proposed vision-based pressure estimation approaches. \citeauthor{PressureVision} \cite{PressureVision} proposed an RGB camera-based pressure estimation method, while \citeauthor{PressureEye} \cite{PressureEye} developed the \ac{PEye} network, which infers pressure images from RGB inputs using a pixel-wise resampling approach. Further advancements were made by \citeauthor{CleverBodyPressure} \cite{CleverBodyPressure} and \citeauthor{ManavarAttFnet} \cite{ManavarAttFnet}, who predicted pressure distributions from single depth images.

\citeauthor{CleverBodyPressure} \cite{CleverBodyPressure} proposed BPBnet and BPWnet to infer contact pressure distributions using a network that embeds a human body mesh model and employs a white-box model for pressure reconstruction. Later, \citeauthor{ManavarAttFnet} \cite{ManavarAttFnet} introduced the \ac{AttnFnet} model, which incorporates mixed-domain loss and a \ac{cGAN} training strategy for pressure inference using the \ac{SLP} dataset.

Building on these prior concepts, the proposed study introduces a novel approach for inferring plausible pressure distributions through conditional generative modeling. This study proposes the inclusion of anthropometric information to build interdependencies via \ac{ILS}, and further improving pressure plausibility through \ac{WOL} when predicting distributions in pixel space. Furthermore, in our work, we synthesized pressure distributions using conditional \ac{BBDM}, evaluated the impact of \ac{ILS} during diffusion denoising \ac{BBDM} + \ac{ILS}, and introduced a two-stage training approach for a conditional \ac{LBBDM} tailored for pressure synthesis. To the authors' knowledge, no prior work has applied diffusion-based modeling for predicting pressure distributions.

\textbf{Optimization for Plausibility:} Although previous methods successfully infer pressure from single depth images, the synthesized pressure distributions often lack physical plausibility, even when achieving low metric scores in \ac{MSE} and \ac{SSIM} \cite{ManavarAttFnet}. While BPWnet and \ac{AttnFnet} outperform earlier baselines in \ac{MSE}, they still fail to produce distributions that align with actual weight measurements \cite{ManavarAttFnet, CleverBodyPressure}. Relying on auxiliary networks like Betanet \cite{CleverBodyPressure} is insufficient, as it only estimates body parameters rather than enforcing physical plausibility directly within the pressure generation pipeline.

Although Betanet predicts height and mass, this component remains independent of BPBnet and BPWnet, and does not establish any interrelationships between pressure and anthropometric attributes. Similarly, the approach proposed by \citeauthor{ManavarAttFnet} \cite{ManavarAttFnet} lacks a mechanism to model such dependencies.

This study addresses these challenges by proposing an informed mechanism to incorporate plausibility into pressure inference through \ac{ILS} and \ac{WOL}. Additionally, \ac{ILS} is integrated into the denoising process of the \ac{BBDM} during training and into the pretrained network during \ac{LBBDM} training to investigate its influence on pressure plausibility in diffusion denoising.

% Summing pressure values and multiplying by the contact area provides an estimate of total body weight, which often deviates from true measurements.

%% file: sec/3_methods.tex
\section{Methods}
\label{methods}

\begin{figure}[h]
  \centering
  \includegraphics[width=0.99\linewidth]{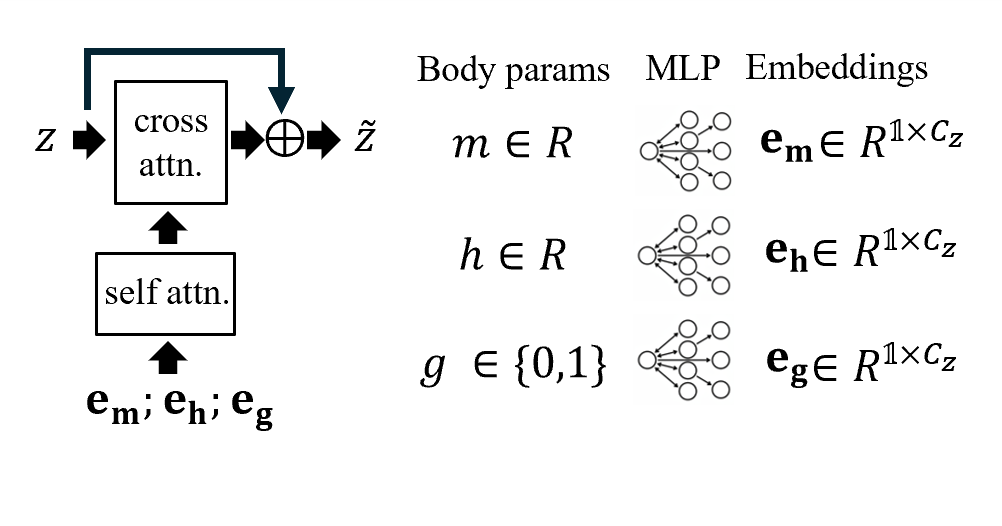}
  \caption{Overview of the ILS module, consisting of anthropometric embedding self-attention and image-embedding cross-attention to generate the informed latent $\tilde{z}$ from latent $z$.}
  \label{fig:ILS_module}
\end{figure}

The method section describes the inclusion of \ac{ILS}, \ac{WOL}, and pressure synthesis using conditional \ac{BBDM} and \ac{LBBDM} \cite{Li_2023_BBDM}.

\subsection{Drawing Meaningful Samples through ILS}

Let $\mathbf{d} \in \mathbb{R}^{H_d \times W_d}$ represent the input depth image and $\mathbf{p} \in \mathbb{R}^{H_p \times W_p}$ represent the pressure image, where $H_d$ and $W_d$ denote the height and width of the depth image, and $H_p$ and $W_p$ denote the height and width of the pressure image.

\begin{equation}
\label{eq:encoder}
z = \mathcal{E}_\phi(\mathbf{d})
\end{equation}

The depth image is encoded using $\mathcal{E}_\phi(\mathbf{d})$, parameterized by $\phi$. It produces a latent feature map $z \in \mathbb{R}^{C_z \times H_z \times W_z}$ for one sample, as shown in \cref{eq:encoder}, where $C_z$ represents the latent embedding dimension.

%. For a batch, $z \in \mathbb{R}^{B \times C_z \times H_z \times W_z}$

\begin{equation}
\label{eq:decoder}
\mathbf{\hat{p}} = \mathcal{D}_\theta\left(z ;\left\{\mathbf{s}_1, \mathbf{s}_2, \ldots, \mathbf{s}_L\right\}\right)
\end{equation}

The decoder $\mathcal{D}_\theta\left(z; \left\{\mathbf{s}_1, \mathbf{s}_2, \ldots, \mathbf{s}_L\right\}\right)$ takes the latent feature map along with optional skip connections and produces the pressure image $\mathbf{\hat{p}}$.

The latent feature map $z$ produced by the encoder has no \ac{BM}, body-height, or gender awareness during training if the model is trained without incorporating additional information into the model architecture. To encode anthropometric information into the model, we condition this information in the latent space.

Let the anthropometric information include \ac{BM} $m \in \mathbb{R}$, height $h \in \mathbb{R}$, and gender $g \in \{0,1\}$. These parameters pass through a learnable linear layer to generate encodings, as shown in \cref{eq: anthropometric_emb}. Before passing the \ac{BM} $m$ and body-height $h$ values to the MLP layer, we normalize them by dividing by the largest possible values of \ac{BM} and body-height in the dataset.

\begin{equation}
\begin{aligned}
& \mathbf{e}_m=\operatorname{MLP}_m\left(m ; \theta_m\right) \in \mathbb{R}^{1 \times C_z} \\
& \mathbf{e}_h=\operatorname{MLP}_h\left(h ; \theta_h\right) \in \mathbb{R}^{1 \times C_z} \\
& \mathbf{e}_g=\operatorname{MLP}_g\left(g ; \theta_g\right) \in \mathbb{R}^{1 \times C_z}
\end{aligned}
\label{eq: anthropometric_emb}
\end{equation}

The individual parameter embeddings are aggregated into a unified conditioning representation resulting in a sequence of three conditioning tokens \cref{eq: concat}, where $[\cdot ; \cdot]$ denotes concatenation.

\begin{equation}
\begin{aligned}
&\mathbf{E_{m,h,g}}=\left[\mathbf{e}_m ; \mathbf{e}_h ; \mathbf{e}_g\right] \in \mathbb{R}^{3 \times C_z}\\
\end{aligned}
\label{eq: concat}
\end{equation}

Concatenated encodings then pass through standard multi-headed self-attention layer to model inter-parameter relationships, as shown in \cref{eq: MHA_mhg} \cite{NIPS2017_Attention}. 

\begin{equation}
\operatorname{Attn}_{m, h, g}= layernorm(\operatorname{MHA}\left(E_{m, h, g}, E_{m, h, g}, E_{m, h, g}\right))
\label{eq: MHA_mhg}
\end{equation}

% where $MHA(\cdot)$ denotes the standard multi-head self-attention from Attention Is All You Need \cite{NIPS2017_Attention}.

To condition the anthropometric attention output \cref{eq: MHA_mhg}, the latent representation $z$ is first flattened along the sequence dimensions and transposed for cross-attention conditioning: $z \in \mathbb{R}^{d_z \times C_z}$, where $d_z$ denotes the sequence length ($H_z \times W_z$).

Finally, to integrate anthropometric information into the latent space, the attention output of the anthropometric information $\operatorname{Attn}_{m,h,g}$ \cref{eq: MHA_mhg} is used as the key in a multi-headed cross-attention operation on the latent representation $z$, as illustrated in \cref{eq: img_tok_attn}.

\begin{equation}
\tilde{z}= layernorm\left(z + \operatorname{MHA}\left(z, \operatorname{Attn}_{m, h, g}, \operatorname{Attn}_{m, h, g}\right)\right)
\label{eq: img_tok_attn}
\end{equation}

Instead of drawing samples from the latent representation $z$, the decoder now draws more meaningful samples from the informed latent $\tilde{z} \in \mathbb{R}^{C_z \times H_z \times W_z}$, improving the plausibility of the generated pressure distributions \cref{eq: decode_informed}.
 
\begin{equation}
\tilde{\mathbf{p}} = \mathcal{D}_\theta\left(\tilde{z} ;\left\{\mathbf{s}_1, \mathbf{s}_2, \ldots, \mathbf{s}_L\right\}\right)
\label{eq: decode_informed}
\end{equation}

\subsection{Reducing Total Body-Mass Error through WOL}
\label{sec:wol}

The recorded actual pressures are obtained from the pressure mat positioned beneath the subjects \cite{LiuSLP}. For bodies at rest, the force acting upon the pressure mat corresponds to the gravitational force $(Mg)$, where $M$ denotes the \ac{BM} of the person lying on the bed, and $g$ represents the gravitational acceleration.

The pressure acting on the $i^{th}$ sensor is defined by \cref{eq: pressure}.

\begin{equation}
p_i = \frac{M_i g}{A_i}
\label{eq: pressure}
\end{equation}

The total \ac{BM} of the person can be estimated from the pressure distribution using \cref{eq:mass} \cite{ManavarAttFnet}, where $N$ is the total number of sensors and $A_i$ denotes the area of the $i^{th}$ sensor.

\begin{equation}
M = \sum_{i=1}^N \frac{p_{i} A_i}{g}
\label{eq:mass}
\end{equation}

The predicted \ac{BM} $\hat{M}$ is computed from the predicted pressure distribution $\hat{p}$ using \cref{eq:mass}. To minimize the body-mass estimation error, the mean absolute difference between the predicted \ac{BM} $\hat{M}$ and the calculated \ac{BM} $M$ is used. From \cref{eq:mass} we can derive \ac{WOL} as \cref{eq:wol} (more in supplimentary material \cref{sec:wol_derivation}), which reduces the total \ac{BM} error and improves the plausibility of the pressure distributions inferred from the depth image.

\begin{equation}
\mathcal{L}_{WOL}=  |\sum_{i=1}^N{(p_{i} - \hat{p_{i}})}|  %\frac{A}{g}
\label{eq:wol}
\end{equation}

% \begin{equation}
% \mathcal{L}_{WOL}=  \sum_{i=1}^N {\|P_{i} - \hat{P_{i}}}\|  %\frac{A}{g}
% \label{eq:wol_inequal}
% \end{equation}

\subsection{Training Strategy}
\label{sec:training_strategy}

This study employs conditional generative modeling approaches such as \ac{cGAN}, \ac{BBDM}, and \ac{LBBDM} to synthesize pressure distributions.

We have used the encoder-decoder model architecture introduced by \citeauthor{ManavarAttFnet} \cite{ManavarAttFnet} when predicting pixels. For diffusion model training, we use OpenAI's improved U-Net as the denoising model, consistent with the setup by \citeauthor{Li_2023_BBDM} \cite{Li_2023_BBDM}.

As our first use case, the \ac{AttnFnet} model is trained with a combination of adversarial loss, perceptual loss, pixel loss, and weight optimization loss (WOL). The model is trained with the objectives shown in \cref{eq: gan_loss}, \cref{eq: generator_loss}, additionaly incorporating the \ac{ILS}.

\begin{equation}
\begin{aligned}
  \mathcal{L}_D = -\Big[{} &
    \mathbb{E}_{\mathbf{d},\mathbf{p}} \big[ y_{\text{real}} \cdot \log(D(\mathbf{d}|\mathbf{p})) \big] \\
  &+ \mathbb{E}_{\mathbf{d},\mathbf{p}} \big[ (1-y_{\text{real}}) \cdot \log(1 - D(\mathbf{d}|\mathbf{p})) \big] \\
  &+ \mathbb{E}_{\mathbf{d}} \big[ y_{\text{gen}} \cdot \log(D(\mathbf{d}|G(\mathbf{d}))) \big] \\
  &+ \mathbb{E}_{\mathbf{d}} \big[ (1-y_{\text{gen}}) \cdot \log(1 - D(\mathbf{d}|G(\mathbf{d}))) \big]
  \Big]
\end{aligned}
\label{eq: gan_loss}
\end{equation}

\begin{equation}
\begin{aligned}
  \mathcal{L}_G = \Big[{} & -\mathbb{E}_{\mathbf{d}}[\log(D(\mathbf{d}|G(\mathbf{d}))))] \\
  &+ \lambda \cdot \mathbb{E}_{\mathbf{d},\mathbf{p}} [\alpha \mathcal{L}_{SSIM} + \beta \mathcal{L}_{L2} + \gamma \mathcal{L}_{WOL}] 
  \Big]
  \label{eq: generator_loss}
\end{aligned}
\end{equation}

The discriminator is a PatchGAN like \cite{pix2pix2017}. Label values for $y_{gen}$ and $y_{real}$ are set to 0.1 and 0.9, respectively. The loss weight for $\lambda$ is set to 100, and the ratio of $\alpha$, $\beta$, and $\gamma$ is $3:0.01:0.01$. Here, $\mathcal{L}_{SSIM}$ is the \ac{SSIM} loss, and $\mathcal{L}_{L2}$ is the \ac{MSE} loss \cite{ManavarAttFnet}.

For the \ac{BBDM} training, we follow the procedure described by \citeauthor{Li_2023_BBDM} \cite{Li_2023_BBDM} using OpenAI's U-Net model. However, the condition $\boldsymbol{y}$ is concatenated with the diffused image $\boldsymbol{x}_t$ as input to the model, enabling it to retain the conditioning information. The following objective is used during training \cref{eq: BBDM_obj}.

\begin{equation}
\mathbb{E}_{\boldsymbol{x}_0,\boldsymbol{y},\boldsymbol{\epsilon}}[||m_t(\boldsymbol{y}-\boldsymbol{x}_0)+\sqrt{\delta_t}\boldsymbol{\epsilon}-\boldsymbol{\epsilon}_\theta(\boldsymbol{x}_t|\boldsymbol{y},t)||^2]
\label{eq: BBDM_obj}
\end{equation}

Here, $\boldsymbol{y}$ is the conditioning input (depth image $\mathbf{d}$), $\boldsymbol{x}_0$ is the clean target image (pressure image $\mathbf{p}$), and $\boldsymbol{\epsilon} \in \mathcal{N}(0,I)$ represents Gaussian noise. The variable $\boldsymbol{x}_t$ denotes the diffused data (diffused pressure image) at time step $t$, $\boldsymbol{\epsilon}_\theta$ is the denoising function, and $m_t$ is defined as $m_t = \frac{t}{T}$, where $T$ denotes the total number of diffusion time steps. The term $\delta_t$ represents the variance.

The forward diffusion process is described in \cref{eq: forward_diff}.

\begin{equation}
\boldsymbol{x}_t=(1-m_t)\boldsymbol{x}_0+m_t\boldsymbol{y}+\sqrt{\delta_t}\boldsymbol{\epsilon}_t
\label{eq: forward_diff}
\end{equation}

To infer the pressure image $\mathbf{p}$ from the depth image $\mathbf{d}$, we employ a non-Markovian process while preserving the original marginal distribution as in a Markovian inference process. We use a sequence length of $S = 200$ to accelerate the sampling process. For the diffusion steps $t = [0:T]$, a marginally distributed subset is defined by $t_s = [0, \tau_1, \tau_2, ... , \tau_s]$.

% \begin{table}[h]
% \captionlistentry{Sampling hyperparameters}
% \label{tab: sampling}
%   \centering
%   \begin{tabularx}{\columnwidth}{lXXXX}
%     \toprule
%     \multicolumn{2}{c}{Sampling Algorithm} \\
%     \midrule
%     1. & Sample conditional input $\boldsymbol{x}_T = y \sim  q(\mathbf{d})$ \\
%     2. & \textbf{for} $ t_s = \tau_s, ..., \tau_1 $ \textbf{do}  \\
%     3. & \quad $z\sim\mathcal{N}(0,\mathbf{I})\mathrm{~if~}t_s>1,\mathrm{else~}z=0$  \\
%     4. & \quad $ \boldsymbol{x}_{t_s-1} = (1-m_{t_s-1}) \boldsymbol{x}_0 + m_{t_s-1} \boldsymbol{y} + $\\
%        & \quad $\sqrt{\delta_{t_s-1}-\sigma_{t_s}^2}\frac{1}{\sqrt{\delta_{t_s}}} \left(x_{t_s}-(1-m_{t_s})x_0-m_{t_s}y\right) + $ \\
%        & \quad $\sqrt{\tilde{\delta}_t}z$ \\
%     \bottomrule
%   \end{tabularx}
% \end{table}

% \begin{equation}
% \begin{aligned}
%   \boldsymbol{x}_{t_s-1} = \Big[{} &(1-m_{t_s-1}) \boldsymbol{x}_{0} + m_{t_s-1} \boldsymbol{y} + \\
%   &\sqrt{\delta_{t_s-1}-\tilde{\delta}_t}\frac{1}{\sqrt{\delta_{t_s}}} \left(x_{t_s}-(1-m_{t_s})x_0 - m_{t_s}y\right) + \\
%   &\sqrt{\tilde{\delta}_t} z
%   \Big]
%   \label{eq: sampling_step}
% \end{aligned}
% \end{equation}

\begin{equation}
\label{eq: sampling_step}
\begin{aligned}
\boldsymbol{x}_{t_s-1}
  ={}& (1-m_{t_s-1})\,\boldsymbol{x}_{\hat{0}} + m_{t_s-1}\,\boldsymbol{y} \\
  & {} + \sqrt{\frac{\delta_{t_s-1}-\tilde{\delta}_{t_s}}
        {\delta_{t_s}}}\,
        \big(\boldsymbol{x}_{t_s}-(1-m_{t_s})\boldsymbol{x}_{\hat{0}} - m_{t_s}\boldsymbol{y}\big) \\
  & {} + \sqrt{\tilde{\delta}_{t_s}}\, z
\end{aligned}
\end{equation}

\begin{table}[h]
\captionlistentry{Sampling hyperparameters}
\label{tab: sampling}
  \centering
  \begin{tabularx}{\columnwidth}{lX}
    \toprule
    \multicolumn{2}{c}{Sampling Algorithm} \\
    \midrule
    1. & Sample conditional input $\boldsymbol{x}_T = \boldsymbol{y} \sim  q(\mathbf{d})$ \\
    2. & \textbf{for} $ t_s = \tau_s, ..., \tau_1 $ \textbf{do}  \\
    3. & \quad $z\sim\mathcal{N}(0,\mathbf{I})\mathrm{~if~}t_s>1,\mathrm{else~}z=0$  \\
    4. & \quad Sample $\boldsymbol{x}_{t_s-1}$ \cref{eq: sampling_step} \\
    5. & \textbf{return} $\boldsymbol{x}_0$ \\
    \bottomrule
  \end{tabularx}
\end{table}

During the \ac{BBDM} forward process, the data $\boldsymbol{x}_0$ gradually converges toward $\boldsymbol{y}$. By doing so, the model is encouraged to predict noise in a manner that reconstructs the original pressure distribution from the conditioning input $\boldsymbol{y}$ (depth image) during inference. Algorithm \ref{tab: sampling} describes the sampling procedure of \ac{BBDM}, where $\boldsymbol{x}_{\hat{0}}$ is defined as $\boldsymbol{x}_{t_s} - \boldsymbol{\epsilon}_\theta(\boldsymbol{x}_t|\boldsymbol{y},t)$, and $\tilde{\delta}_{t_s}$ represents the posterior variance.

Because \ac{BBDM} predicts noise, we cannot use \ac{WOL} \cref{eq:wol} since it is irrelevant to the main objective. However, we integrate \ac{ILS} into OpenAi's U-Net architecture to evaluate whether it can help the denoising process and produce more plausible results.

Finally, we employed \ac{BBDM} in the latent space to evaluate computational performance improvements alongside the quality of generated pressure distributions. The \ac{AttnFnet} model was trained in a self-reconstruction manner using \ac{ILS} to encode the relationships among mass, height, and gender during image reconstruction. Skip connections in the \ac{AttnFnet} architecture were removed to minimize the influence of the encoder during the decoding process. To reduce the size of the denoising model, a second bottleneck layer was introduced after the bottleneck layer to decrease the feature channel dimensionality (see supplementary \cref{fig:lbbdm_strategy}).

The model was trained in an unconditional \ac{GAN} setting with an unconditional PatchGAN discriminator and a modified \ac{AttnFnet}. The following objectives were used during pretraining \cref{eq: unconditional_gan_loss} and \cref{eq: reconstruction_loss}, where $\mathbf{x}$ denotes the model input and $\mathbf{\hat{x}}$ represents the reconstructed output generated by the autoencoder.

% \begin{equation}
% \begin{aligned}
%   \mathcal{L}_D = -\Big[{} &
%     \mathbb{E}_{\mathbf{x}} \big[ y_{\text{real}} \cdot \log(D(\mathbf{x})) \big] \\
%   &+ \mathbb{E}_{\mathbf{x}} \big[ (1-y_{\text{real}}) \cdot \log(1 - D(\mathbf{x})) \big] \\
%   &+ \mathbb{E}_{\mathbf{x}} \big[ y_{\text{gen}} \cdot \log(D(G(\mathbf{x}))) \big] \\
%   &+ \mathbb{E}_{\mathbf{x}} \big[ (1-y_{\text{gen}}) \cdot \log(1 - D(G(\mathbf{x}))) \big]
%   \Big]
% \end{aligned}
% \label{eq: unconditional gan loss}
% \end{equation}

\begin{equation}
\begin{aligned}
  \mathcal{L}_D = 
    - \mathbb{E}_{\mathbf{x}} \Big[ {} & y_{\text{real}} \cdot \log(D(\mathbf{x}))  \\
  &+  (1-y_{\text{real}}) \cdot \log(1 - D(\mathbf{x}))  \\
  &+  y_{\text{gen}} \cdot \log(D(G(\mathbf{x})))  \\
  &+ (1-y_{\text{gen}}) \cdot \log(1 - D(G(\mathbf{x}))) 
  \Big]
\end{aligned}
\label{eq: unconditional_gan_loss}
\end{equation}

\begin{equation}
\begin{aligned}
  \mathcal{L}_G = \Big[{} & -\mathbb{E}_{\mathbf{x}}[\log(D(G(\mathbf{x}))))] \\
  &+ \lambda \cdot \mathbb{E}_{\mathbf{x},\mathbf{\hat{x}}} [\alpha \mathcal{L}_{SSIM} + \beta \mathcal{L}_{L2}] 
  \Big]
  \label{eq: reconstruction_loss}
\end{aligned}
\end{equation}

Post-training was performed using \ac{LBBDM} in the latent space, following the objective described in \cref{eq: BBDM_obj}. Samples were inferred from the latent space of the depth image using Algorithm \ref{tab: sampling} to obtain the pressure latent representation. The \ac{AttnFnet} decoder was subsequently used to generate pressure distributions from the denoised latent features.

This study also utilized an unconditionally trained VQGAN \cite{VQGAN2022} model on the CelebHQ dataset \cite{CelebHQ2020}, in a manner consistent with the \ac{BBDM} paper \cite{Li_2023_BBDM} and compared results.

%% file: sec/4_evaluation.tex
\section{Evaluation}
\label{evaluation}

\begin{table}[h]
  \addtocounter{table}{-1}
  \centering
  \begin{tabularx}{\columnwidth}{lYYYY}
    \toprule
    \multicolumn{5}{c}{Real + Synthetic Data ($208K$)} \\ \cline{1-5} 
    Img Description & Ucov & cov \,($1$mm) & cov ($3$mm) & total Img\\
    \midrule
    real train (61) & 2745 & 2745 & 2745 & 8235 \\
    real val (20) & 900 & 900 & 900 & 2700 \\
    real test (20) & 900 & 900 & 900 & 2700 \\
    total real & 4500 & 4500 & 4500 & 13635 \\ \cline{1-5} 
    Synth Img & 97495 & \multicolumn{2}{c}{97495} & 194990 \\ \cline{1-5}
    Total training & & & & 203225 \\
    Total validation & & & & 2700 \\
    Total test & & & & 2700 \\

    \bottomrule
  \end{tabularx}
  \caption{Dataset splits and counts of uncovered (Ucov) and covered (cov) images, including $1$ mm and $3$ mm coverage variants.}
  \label{tab:dataset}
\end{table}

\begin{table*}[ht]
    \centering
    % \begin{tabular}{|p{2.5cm}|p{1.2cm}|p{1.2cm}|p{1.5cm}|p{1.5cm}|p{1.5cm}|p{1.5cm}|}
    \begin{tabular}{@{}lclclclc@{}}
        \toprule
        Method & Model & MPPA & MSSIM & MFID & MSE & MPSNR \\
        \midrule
        \multicolumn{7}{c}{$3\,\mathrm{k}$ training samples} \\ [0.2em] \cline{1-7} \\ [-0.9em]

        \citeauthor{ManavarAttFnet} \cite{ManavarAttFnet} & U-Net &  0.6658 & 0.7958 & 0.4615 & 0.000433 & 34.4185 \\
        
        \citeauthor{ManavarAttFnet} \cite{ManavarAttFnet} & \acs{AttnFnet}-143M & 0.6142 & 0.8291 & 0.3475 & 0.000368  &  35.3508 \\

        \citeauthor{CleverBodyPressure} \cite{CleverBodyPressure} & BPBnet  & 0.0078 & 0.0204 & 160.58 & 0.00567 & 22.5927 \\

        \citeauthor{CleverBodyPressure} \cite{CleverBodyPressure} & BPWnet & 0.5244 & 0.6331 & 1.6340 & 0.00405 & 24.1364 \\

        % Ours & U-Net & - & - & - & - & - \\

        Ours & \ac{AttnFnet}-11M & 0.6412 & \textbf{0.8864} & 1.0946 & 0.000263 & 36.2673 \\

        BBDM & Openai U-Net & 0.0599 & 0.7783 & \textbf{0.1977} & \textbf{0.000197} & \textbf{37.3513} \\

        BBDM + ILS & Openai U-Net & 0.5992 & 0.8730 & 0.6197 & 0.000346 & 34.9381 \\

        LBBDM VQGAN & pretrained VQGAN & 0.0092 & 0.2611 & 1.5429 & 0.000777 & 31.2064 \\
        
        LBBDM & proposed pretraining & \textbf{0.6950} & 0.8510 & 0.7338 & 0.000349 & 34.93 \\ [0.2em] \cline{1-7} \\ [-0.9em]
        
        % LBBDM + L1 loss & proposed pretraining & \textbf{0.6885} & 0.8687 & 1.0123 & 0.000309 & 35.5299 \\ [0.2em] \cline{1-7} \\ [-0.9em]

        \multicolumn{7}{c}{\centering $8\,\mathrm{k}$ training samples} \\ [0.2em] \cline{1-7} \\ [-0.9em]
        
        \citeauthor{ManavarAttFnet} \cite{ManavarAttFnet} & \acs{AttnFnet}-11M & 0.6793 & 0.8847 & 0.4323 & 0.000264 & 36.3225 \\ 

        Ours & \ac{AttnFnet}-11M & 0.6451 & 0.8795 & 0.6979 & 0.000270 & 36.1675 \\

        BBDM & Openai U-Net & 0.6772 & 0.9142 & 0.4381 & 0.000269 & 36.2827 \\

        BBDM + ILS & Openai U-Net & 0.6679 & \textbf{0.9158} & \textbf{0.3465} & \textbf{0.000259} & \textbf{36.5109} \\

        LBBDM & proposed pretraining & \textbf{0.7068} & 0.8507 & 0.9406 & 0.000313 & 35.4798 \\ [0.2em] \cline{1-7} \\ [-0.9em]
        
        % LBBDM + L1 loss & proposed pretraining & \textbf{0.7017} & 0.8536 & 0.7269 & 0.000319 & 35.3887 \\ [0.2em] \cline{1-7} \\ [-0.9em]

        \multicolumn{7}{c}{\centering $203\,\mathrm{k}$ training samples} \\ [0.2em] \cline{1-7} \\ [-0.9em]

        \citeauthor{ManavarAttFnet} \cite{ManavarAttFnet} & \acs{AttnFnet}-11M & \textbf{0.6580} & 0.8860 & 0.8014 & 0.000262 & 36.4592 \\ 

        Ours $\gamma = 0.0182$ & \ac{AttnFnet}-11M & 0.6539 & \textbf{0.8942} & 0.7993 & \textbf{0.000247} & \textbf{36.7770} \\

        BBDM & Openai U-Net & 0.6306 & 0.8901 & 0.4267 & 0.000361 & 34.9244 \\

        BBDM + ILS & Openai U-Net & 0.4146 & 0.8630 & \textbf{0.3821} & 0.000431 & 34.18 \\

        \bottomrule
    \end{tabular}
  \caption{Test‑set comparison of MPPA, MSSIM, MFID, MSE, and MPSNR across methods at three training sample sizes. \acs{AttnFnet}-143M and \acs{AttnFnet}-11M describes \acs{AttnFnet} model with 143 million and 11 million parameters respectively.}
  \label{tab:metric_scores}
\end{table*}

% All experiments were conducted using the dataset proposed by \citeauthor{LiuSLP} \cite{LiuSLP}, and the synthetic dataset by \citeauthor{DVN_Clever_Synthetic} \cite{DVN_Clever_Synthetic}. The proposed methods were compared with baseline methods from \citeauthor{CleverBodyPressure} \cite{CleverBodyPressure} and \citeauthor{ManavarAttFnet} \cite{ManavarAttFnet}.

\subsection{Dataset}

The \acf{SLP} dataset \cite{LiuSLP} contains 102 healthy participants under three cover conditions and 45 unique poses, grouped into three sleeping postures (supine, left lateral, and right lateral). Cover conditions include uncovered poses, poses covered by a 1 mm blanket, and poses covered by a 3 mm blanket.

All experiments used both depth and pressure image modalities, with anthropometric information (\ac{BM}, body-height, gender) integrated. \ac{OFDI} data were used in place of raw depth images \cite{DVN_Clever, ManavarAttFnet}. Synthetic samples generated by \citeauthor{DVN_Clever_Synthetic} \cite{DVN_Clever_Synthetic} augmented training data, including both uncovered and covered poses representing the real data blanket conditions \cite{CleverBodyPressure}.

Experiments followed a $60:20:20$ split for training, validation, and testing. The splits were based on the 102 real participants, while synthetic data were added only to training. Subject 7 was excluded due to noise cleaning errors in the depth data \cite{DVN_Clever}. \Cref{tab:dataset} shows the detailed splits and image counts.

Real pressure images were pre-processed to reduce ambiguity, minimize data size, and align with prior studies. All real pressure images were resized to $27 \times 64$ and smoothed using a Gaussian filter ($\sigma = 1.4$) as in previous works \cite{CleverBodyPressure, ManavarAttFnet}. The original pressure range for synthetic data was $(0, 101)$kPa and for real data $(0, 111)$kPa; applying $\sigma = 1.4$ eliminated low-frequency high-pressure spikes and pressure in real data ranged between $(0, ~57)$kPa.

Depth arrays were normalized individually to $(0,1)$ and pressure arrays were normalized globally by dividing by the maximum pressure value in the dataset.

\subsection{Data Analysis}

\begin{table}[h]
  \centering
  \begin{tabularx}{\columnwidth}{lXYYX}
    \toprule 
    Method & Ucov & cov ($1$mm) & cov ($3$mm) & overall\\
    \midrule
    \multicolumn{5}{c}{$3\,\mathrm{k}$ training samples} \\ [0.2em] \cline{1-5} \\ [-0.9em]
    \citeauthor{ManavarAttFnet} \cite{ManavarAttFnet} & 0.6884 & - & - & 0.6884 \\
    Ours & 0.7205 & - & - & 0.7205 \\
    BBDM & \textbf{0.4817} & - & - & \textbf{0.4817} \\
    BBDM + ILS & 0.8429 & - & - & 0.8429 \\
    LBBDM VQGAN & 8.0909 & - & - & 8.0909 \\
    LBBDM & 0.9580 & - & - & 0.9580 \\ [0.2em] \cline{1-5} \\ [-0.9em]
    % LBBDM - L1 Loss & 0.8482 & - & - & 0.8482 \\ [0.2em] \cline{1-5} \\ [-0.9em]

    \multicolumn{5}{c}{$8\,\mathrm{k}$ training samples} \\ [0.2em] \cline{1-5} \\ [-0.9em]

    \citeauthor{ManavarAttFnet} \cite{ManavarAttFnet} & 0.7212 & 0.9134 & 0.9363 & 0.8570 \\
    BPWnet \cite{CleverBodyPressure} & 1.470 & 1.444 & 1.455 & 1.456 \\
    Ours & 0.7567 & 0.9291 & 0.9442 & 0.8767 \\
    BBDM & 0.5583 & 0.8146 & 0.8423 & 0.7384 \\
    BBDM + ILS & \textbf{0.5141} & \textbf{0.7994} & \textbf{0.8224} & \textbf{0.7120} \\
    LBBDM & 0.9047 & 1.0590 & 1.0803 & 1.0149 \\ [0.2em] \cline{1-5} \\ [-0.9em]
    % LBBDM - L1 Loss & 0.9220 & 1.0814 & 1.1061 & 1.0365 \\ [0.2em] \cline{1-5} \\ [-0.9em]

    \multicolumn{5}{c}{$203\,\mathrm{k}$ training samples} \\ [0.2em] \cline{1-5} \\ [-0.9em]
    \citeauthor{ManavarAttFnet} \cite{ManavarAttFnet} & 0.7021 & 0.9224 & 0.9285 & 0.8510 \\
    BPBnet \cite{CleverBodyPressure} 101K & 0.772 & 0.858 & 0.884 & 0.838 \\
    BPWnet \cite{CleverBodyPressure} 101K & 1.155 & 1.190 & 1.209 & 1.184 \\
    Ours $\gamma=0.0182$ & \textbf{0.6406} & \textbf{0.8814} & \textbf{0.8819} & \textbf{0.8013} \\
    BBDM & 0.7990 & 1.077 & 1.0979 & 0.9914 \\
    BBDM + ILS & 1.0031 & 1.2541 & 1.2926 & 1.1833 \\
    \bottomrule
  \end{tabularx}
  \caption{Test set \ac{MSE} in $kPa^2$ for uncovered (Ucov), covered at $1$ mm, covered at $3$ mm, and overall poses, evaluated across methods at three training sizes.}
  \label{tab:MSE_pressure}
\end{table}

\begin{figure*}[ht]
  \centering
    \includegraphics[width=\linewidth]{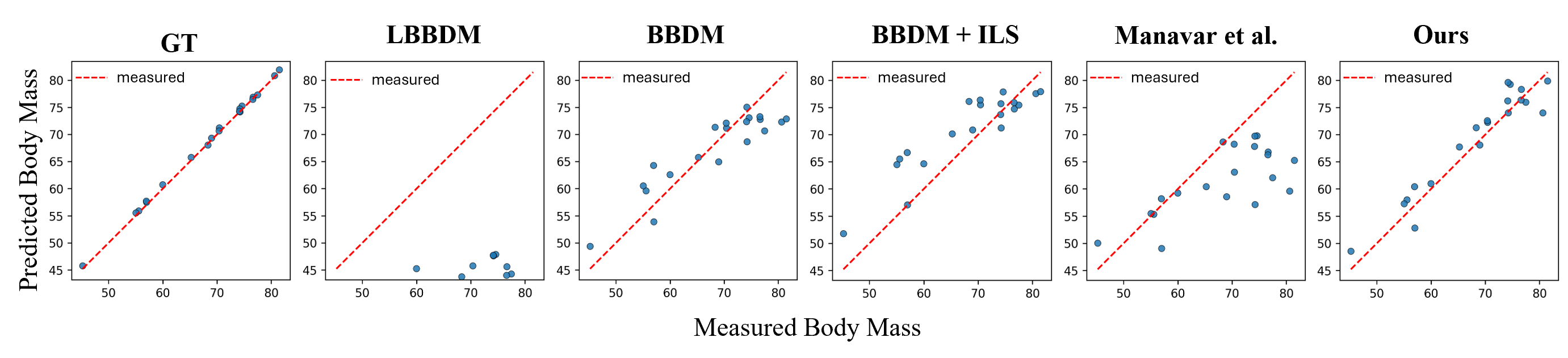}
    \caption{Scatter plot comparing \ac{BM} computed from predicted pressure distributions across methods (blue points) with measured \ac{BM} (red dotted identity line); GT denotes \ac{BM} calculated from ground truth pressure distributions.}
    \label{fig:weights}
  \hfill
\end{figure*}

\begin{figure*}[h]
  \centering
    \includegraphics[width=0.85\linewidth]{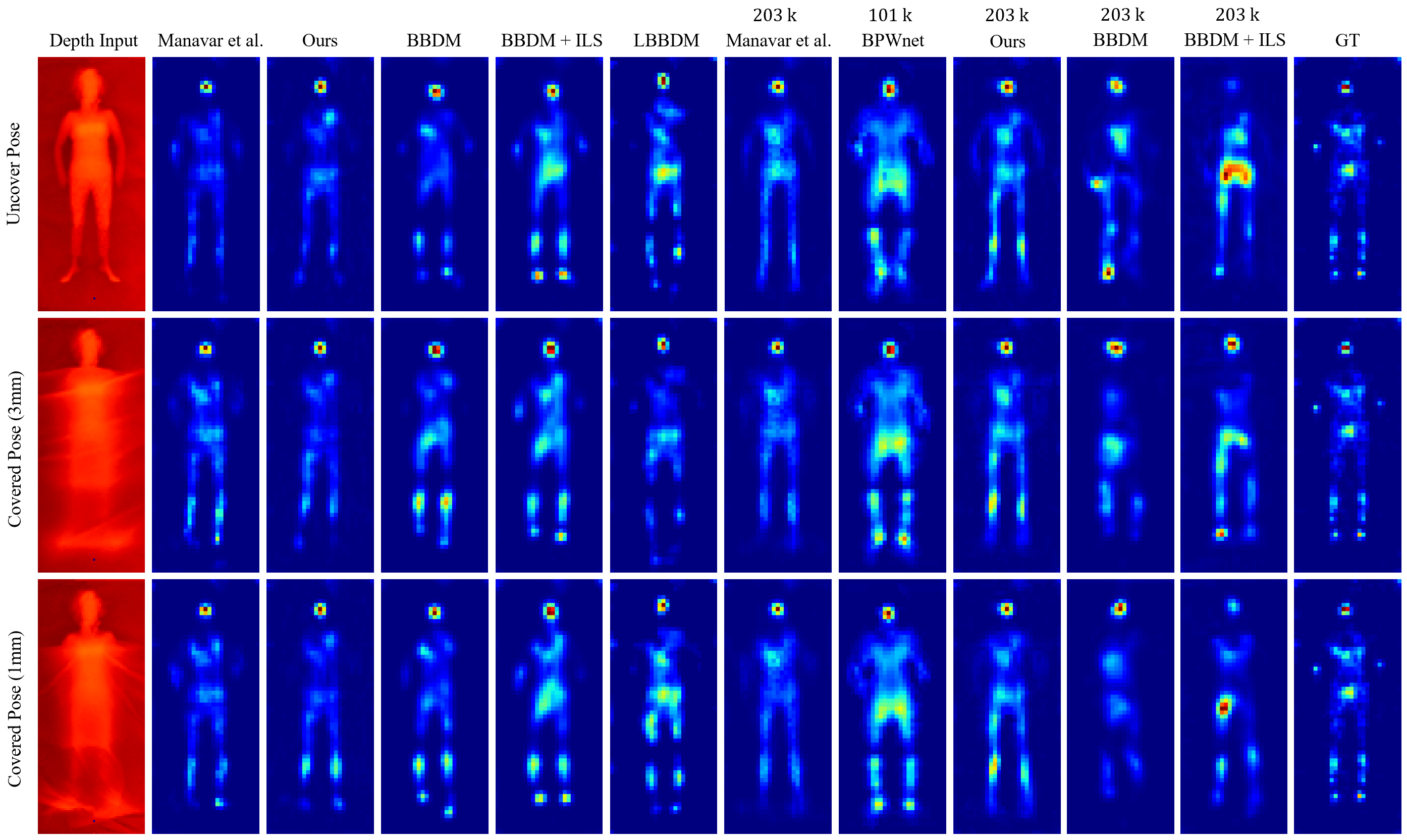}
    \caption{Qualitative comparisons of predicted pressure distributions against ground truths (GT) across representative samples; rows show different inputs to models and, each columns show model outputs.}
    \label{fig:visulization}
  \hfill
\end{figure*}

All models were evaluated under three different training conditions: (1) a sample size of $2745$ ($3\,\mathrm{k}$) using only uncovered postures from real data; (2) a sample size of $8235$ ($8\,\mathrm{k}$) including cover conditions from real data; and (3) a sample size of $203225$ ($203\,\mathrm{k}$) combining both real and synthetic data with all postures and covered conditions. All results are reported at epoch 100, unless stated otherwise.

Standard evaluation metrics for 2D images were used to assess pressure generation quality: \ac{MPPA} \cite{ManavarAttFnet}, \ac{MSSIM} \cite{SSIM}, \ac{MFID} \cite{FID}, \ac{MSE}, and \ac{MPSNR} \cite{PSNR}. \ac{FID} was calculated with the original Inception-V3 model weights \cite{FID_InceptionV3}. \Cref{tab:metric_scores} presents metric scores for the proposed methods and baselines on the test dataset. In \cref{tab:metric_scores}, \ac{BBDM} with \ac{ILS} outperformed all methods when trained on $8\,\mathrm{k}$ samples, while our method with $\gamma = 0.0182$ outperformed in $203\,\mathrm{k}$ training samples.

When images were rescaled from the normalized space, \ac{MSE} was recalculated in pressure space as shown in \cref{tab:MSE_pressure}. Consistent trends were observed: the proposed method displayed increased \ac{MSE} in pressure space, but error rate declined when setting $\gamma=0.0182$. Note that pressure ranges differ between real and synthetic data, and setting $\gamma=1.0$ prevents the model from generating valid samples.

\begin{table}[h]
  \centering
  % \begin{tabular}{|w{c}{1.6cm}|w{c}{1.3cm}|w{c}{1cm}|w{c}{0.5cm}|}
  \begin{tabularx}{\columnwidth}{|c|c|c|}
    \toprule
    Method & \multicolumn{2}{|m{4.8cm}|}{\centering Mean Absolute Body-Mass Difference (kg)} \\
    \cline{2-3}
     &  \multicolumn{1}{|c|}{$M_{\text{meas}} - M_{\text{pred}}$} &  \multicolumn{1}{|c|}{$M_{\text{GT}} - M_{\text{pred}}$}   \\
    \hline
    \citeauthor{ManavarAttFnet} \cite{ManavarAttFnet} &  7.28 &  7.50 \\
    ours &  \textbf{2.56} &  \textbf{2.32} \\
    BBDM &  3.88 &  3.75 \\
    BBDM + ILS &  4.28 &  4.04  \\
    LBBDM & 26.23 & 25.86 \\
    % LBBDM - L1 Loss & 17.36 & 17.73 \\
    BPWnet (Betanet) & 5.64 & - \\
    % BPBnet (Betanet) & 5.64 & - \\
    \bottomrule
  \end{tabularx}
  \caption{Average Mean absolute body mass difference (kg) between measured and ground truth references versus predicted \ac{BM}s, reported as $M_{\text{meas}} - M_{\text{pred}}$ and $M_{\text{GT}} - M_{\text{pred}}$; all models trained on $8\,\mathrm{k}$ real samples. best values are bolded}
  \label{tab:weight_error}
\end{table}

\begin{table}[h]
  \centering
  \begin{tabularx}{\columnwidth}{lXXXX}
    \toprule
    \multicolumn{5}{c}{Posture Intersection Over Union ($203\,\mathrm{k}$ training samples)} \\ %\cline{1-5} 
    \midrule
    Method & Ucov & cov (1mm) & cov (3mm) & overall\\
    \midrule

    \citeauthor{ManavarAttFnet} \cite{ManavarAttFnet} & 0.6561 & 0.6022 & 0.6010 & 0.6198 \\
    Ours $\gamma=0.0182$ & \textbf{0.6675} & \textbf{0.6127} & \textbf{0.6096} & \textbf{0.6299} \\
    BBDM & 0.6285 & 0.5605 & 0.5427 & 0.5772 \\
    BBDM + ILS & 0.5102 & 0.4301 & 0.4214 & 0.4539 \\
    \bottomrule
  \end{tabularx}
  \caption{Posture Intersection over Union (IoU) on the $203\,\mathrm{k}$ training sample training, reported for uncovered (Ucov), covered at $1$ mm, covered at $3$ mm, and overall; best values are bolded.}
  \label{tab:pos_IOU}
\end{table}

Participant \ac{BM}s were calculated from actual and predicted pressure profiles using \cref{eq:mass}. \Cref{tab:weight_error} reports mean absolute \ac{BM} differences among measured \ac{BM}s ($M_{\text{meas}}$), \ac{BM}s calculated from the ground truth pressures ($M_{\text{GT}}$), and from predicted pressures ($M_{\text{pred}}$). Further, \ac{BM} plots in \cref{fig:weights} enables detailed comparisons.

Posture \ac{IOU} \cite{ManavarAttFnet} scores were computed to assess posture prediction error \cref{tab:pos_IOU}, and visual comparisons were used for qualitative analysis \cref{fig:visulization}.

%% file: sec/5_ablation_study.tex
\section{Ablation Study}
\label{ablation}

An ablation study was conducted to assess the impact of \ac{ILS} and \ac{WOL} components on model performance and plausibility of generated results. Diffusion models were also evaluated with varying sampling steps to examine the effects on pressure generation quality.

\Cref{tab:component_impact} demonstrates that combining \ac{ILS} and \ac{WOL} yields more plausible pressure distributions with a slight increase in \ac{MSE}. \Cref{tab:wol_fact} presents the effect of the \ac{WOL} weight $\gamma$ on \ac{MSE} and \ac{BM} \ac{MAE}: reducing $\gamma$ initially reduces the \ac{MSE} but, increases the \ac{BM} error.

For \ac{BBDM}, increasing the number of sampling steps enhances diversity, as reflected by decreasing \ac{FID} scores in \cref{tab:sample_steps}, but also leads to higher \ac{MSE}. The number of sampling steps does not directly affect \ac{BM} estimation accuracy.

\begin{table}[h]
  \centering
  \begin{tabular}{@{}lcc@{}}
    \toprule
    Method  & MSE ($kPa^{2}$) & \ac{BM} MAE\\
    \midrule
    $\ac{ILS}$ & \textbf{0.8601} & 5.70 \\
    $\ac{WOL}$ & 0.9151 & 5.53 \\
    $\ac{ILS} + \ac{WOL}$ & 0.8767 & \textbf{2.56} \\
    \bottomrule
  \end{tabular}
  \caption{Effect of the ILS and WOL components on prediction error: \ac{MSE} of pressure maps and \ac{BM} \acs{MAE}. Model was trained using  $8\,\mathrm{k}$ real samples.}
  \label{tab:component_impact}
\end{table}

\begin{table}[h]
  \centering
  \begin{tabular}{@{}lcc@{}}
    \toprule
    $\gamma$ values  & MSE ($kPa^{2}$) & \ac{BM} MAE\\
    \midrule
    $\gamma=1.0$ & 0.8767 & \textbf{2.56} \\
    $\gamma=0.1$ & 0.8760 & 2.63 \\
    $\gamma=0.0182$ & 0.8903 & 2.74 \\
    $\gamma=0.001$ & \textbf{0.8650} & 4.68 \\
    \bottomrule
  \end{tabular}
  \caption{Effect of the WOL weight $\gamma$ on pressure MSE and \ac{BM} MAE. Model was trained using  $8\,\mathrm{k}$ real samples.}
  \label{tab:wol_fact}
\end{table}

\begin{table}[h]
  \centering
  \begin{tabular}{@{}lccc@{}}
    \toprule
    Sampling steps  & MSE ($kPa^{2}$) & BM MAE & FID\\
    \midrule
    $s=10$ & \textbf{0.6797} & 4.02 & 0.4902\\
    $s=20$ & 0.6900 & 4.07 & 0.4833 \\
    $s=100$ & 0.7273 & \textbf{3.86} & 0.4449 \\
    $s=200$ & 0.7384 & 3.87 & 0.4381 \\
    $s=500$ & 0.7500 & 3.90 & 0.4293 \\
    $s=1000$ & 0.7521 & 3.88 & \textbf{0.4279} \\

    \bottomrule
  \end{tabular}
  \caption{Effect of diffusion sampling steps on pressure \ac{MSE}, \ac{BM} \ac{MAE}, and \ac{FID}; model was trained using  $8\,\mathrm{k}$ real samples.}
  \label{tab:sample_steps}
\end{table}

%% file: sec/6_conclusion.tex
\section{Discussion and Future Work}
\label{discussion_future_work}

As shown in \cref{tab:wol_fact}, decreasing \(\gamma\) reduces \ac{MSE} but weakens \ac{BM}-related information and thus plausibility. A value of \(\gamma=0.0182\) balances these objectives.

According to \cref{tab:metric_scores} and \cref{tab:MSE_pressure}, \ac{BBDM}+ILS and \ac{BBDM} perform best when trained on real data only. However, performance degrades when trained on the larger $203\,\mathrm{k}$ dataset. We speculate that these models require more epochs to converge on the larger dataset compared to our method, though we could not validate this during training due to high inference times and limited compute resources. Nevertheless, \cref{fig:visulization} shows that \ac{BBDM}+ILS produces visually superior results compared to the standard \ac{BBDM}.

A key requirement for using \ac{ILS} in the model is establishing a prior global context for the features, as \ac{ILS} cannot generate valid samples during training without it.

A current limitation is that plausibility is measured only via \ac{BM} error. Future work should include region-specific metrics (e.g., errors at common pressure-injury sites) and additional plausibility criteria.

Another limitation is dataset composition: generation quality reflects the mix of synthetic and limited real data. Future work will expand real datasets, diversify training, and assess cross-domain robustness. Further improvements require including patient data in training and conducting clinical validation, as current experiments involve healthy controls only.

While this study demonstrates promising capabilities for real-time monitoring (see supplementary \cref{sec:realtime_monitoring}), the system currently lacks a comprehensive quantitative assessment and large-scale validation in real-time settings.

\section*{Acknowledgments}
This research was carried out within the framework of the project "SAIL: SustAInable Lifecycle of Intelligent SocioTechnical Systems". SAIL is receiving funding from the program "Netzwerke 2021", an initiative of the Ministry of Culture and Science of the State of North Rhine-Westphalia (grant No.: NW21-059B). The authors would like to acknowledge Dr. Matthias Fricke from the Center for Applied Data Science (CfADS) at Bielefeld University of Applied Sciences for providing access to the GPU compute cluster.

\section*{Conclusion}
\label{Conclusion}

This work predicts physically plausible pressure distributions from a single depth image and explores conditional \ac{BBDM}/\ac{LBBDM}-based pressure synthesis. While \ac{BBDM} with \ac{ILS} yields the strongest metrics, the proposed approach with \ac{ILS} and \ac{WOL} achieves higher plausibility with faster training and inference with only a slight decrease in standard scores. Coupled with \ac{AttnFnet}, it points to a lightweight, non-invasive, real-time patient monitoring system.

%% file: sec/X_suppl.tex
\clearpage
\setcounter{page}{1}
\maketitlesupplementary

\section{Weight Optimation Loss}
\label{sec:wol_derivation}

As shown in \cref{sec:wol} total mass can be describes as \cref{eq:mass2}
\begin{equation}
    m = \sum_{i=1}^{N} \frac{p_i A_i}{g}
    \label{eq:mass2}
\end{equation}

where $p_i$ is the pressure at $i^{th}$ taxel with total $N$ sensors, $A_i$ is the area of the $i^{th}$ taxel, and $g$ is gravitational acceleration.

Assuming predicted pressure at $i^{th}$ taxel is $\hat{p}$ we can get \cref{eq:pred_mass}, having area similar to actual taxel ($A_i = \hat{A}_i$).

\begin{equation}
    \hat{m} = \sum_{i=1}^{N} \frac{\hat{p}_i \hat{A}_i}{g}
    \label{eq:pred_mass}
\end{equation}

To optimize mass consistency, we define the loss as \cref{eq:mass_optim}
\begin{equation}
    L_{\text{WOL}} = | m - \hat{m} | = \left| \sum_{i=1}^{N} \frac{p_i A_i}{g} - \sum_{i=1}^{N} \frac{\hat{p}_i \hat{A}_i}{g} \right|
    \label{eq:mass_optim}
\end{equation}

Since $A_i = \hat{A}_i$ and $\sum_{i=1}^N A_i = A_{\text{tot}}$, we have \cref{eq:wol_2}:

\begin{equation}
    \mathcal{L}_{WOL} = \frac{A_{\text{tot}}}{g} \left| \sum_{i=1}^{N} (p_i - \hat{p}_i) \right|
    \label{eq:wol_2}
\end{equation}

By the triangle inequality, we get \cref{eq:wol_inequal2}.

\begin{equation}
    \mathcal{L}_{WOL} \leq \frac{A_{\text{tot}}}{g} \sum_{i=1}^{N} | p_i - \hat{p}_i |
    \label{eq:wol_inequal2}
\end{equation}

\begin{figure}[t]
  \centering
    \includegraphics[width=\linewidth]{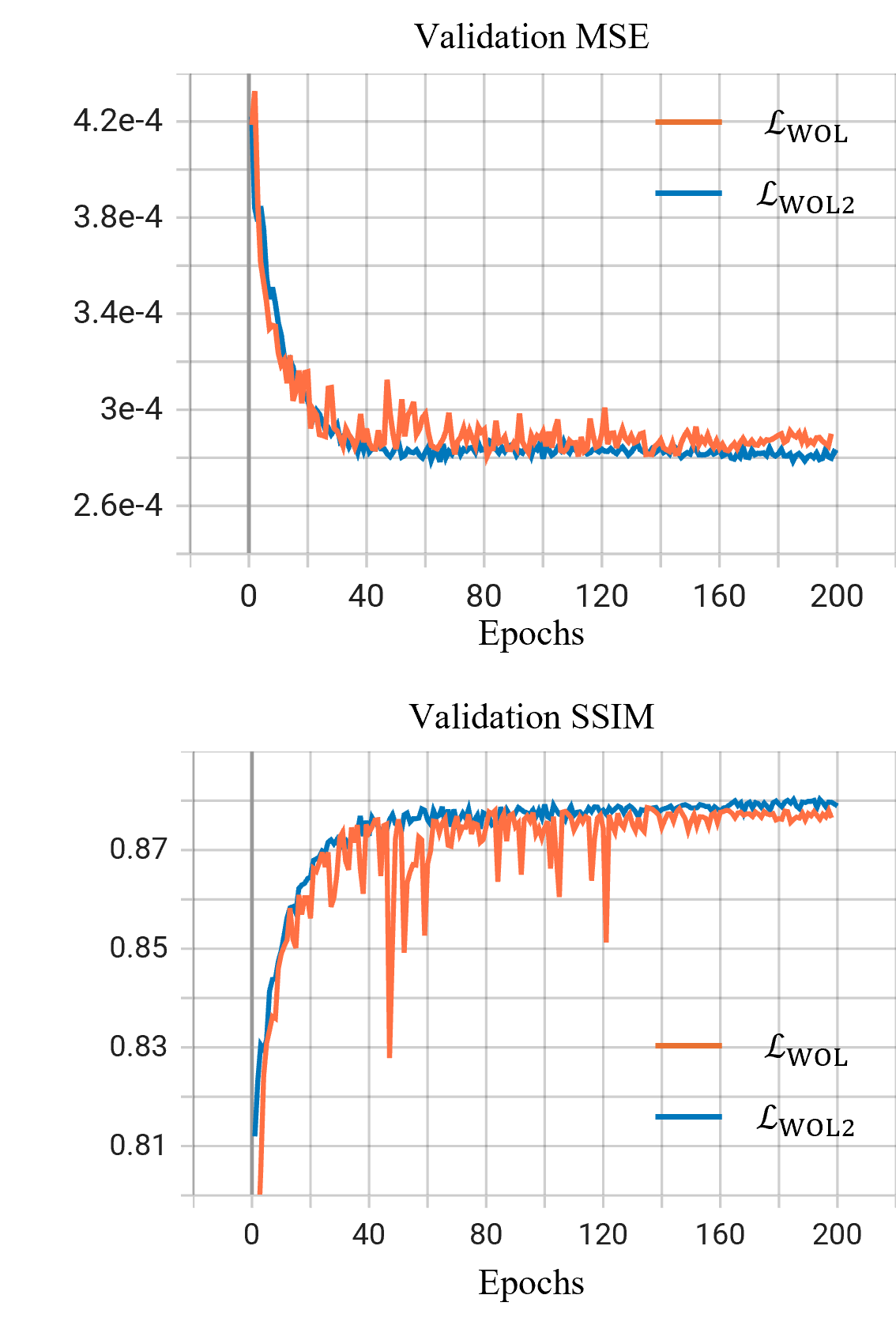}
    \caption{Comparison of validation \ac{MSE} and \ac{SSIM} performance between $\mathcal{L}_{WOL}$ and $\mathcal{L}_{WOL2}$ loss functions. Results are based on models trained for 200 epochs on $8\,\mathrm{k}$ real samples.}
    \label{fig:val_run}
  \hfill
\end{figure}

Therefore, reducing \ac{BM} error is same as minimizing $L_1$ norm \cref{eq:wol_inequal3}. It is more stable (\cref{fig:val_run}) but it can't reduce mass plausibility when low frequency pressure ambiguity in actual distributions:

\begin{equation}
    \mathcal{L}_{WOL2} = \sum_{i=1}^{N} | p_i - \hat{p}_i | = \|\mathbf{p} - \hat{\mathbf{p}} \|_1
    \label{eq:wol_inequal3}
\end{equation}

\section{Effects of Data Normalization and Pre-processing}
\label{sec:eff_data_norm}

\begin{figure*}[h]
  \centering
    \includegraphics[width=0.95\linewidth]{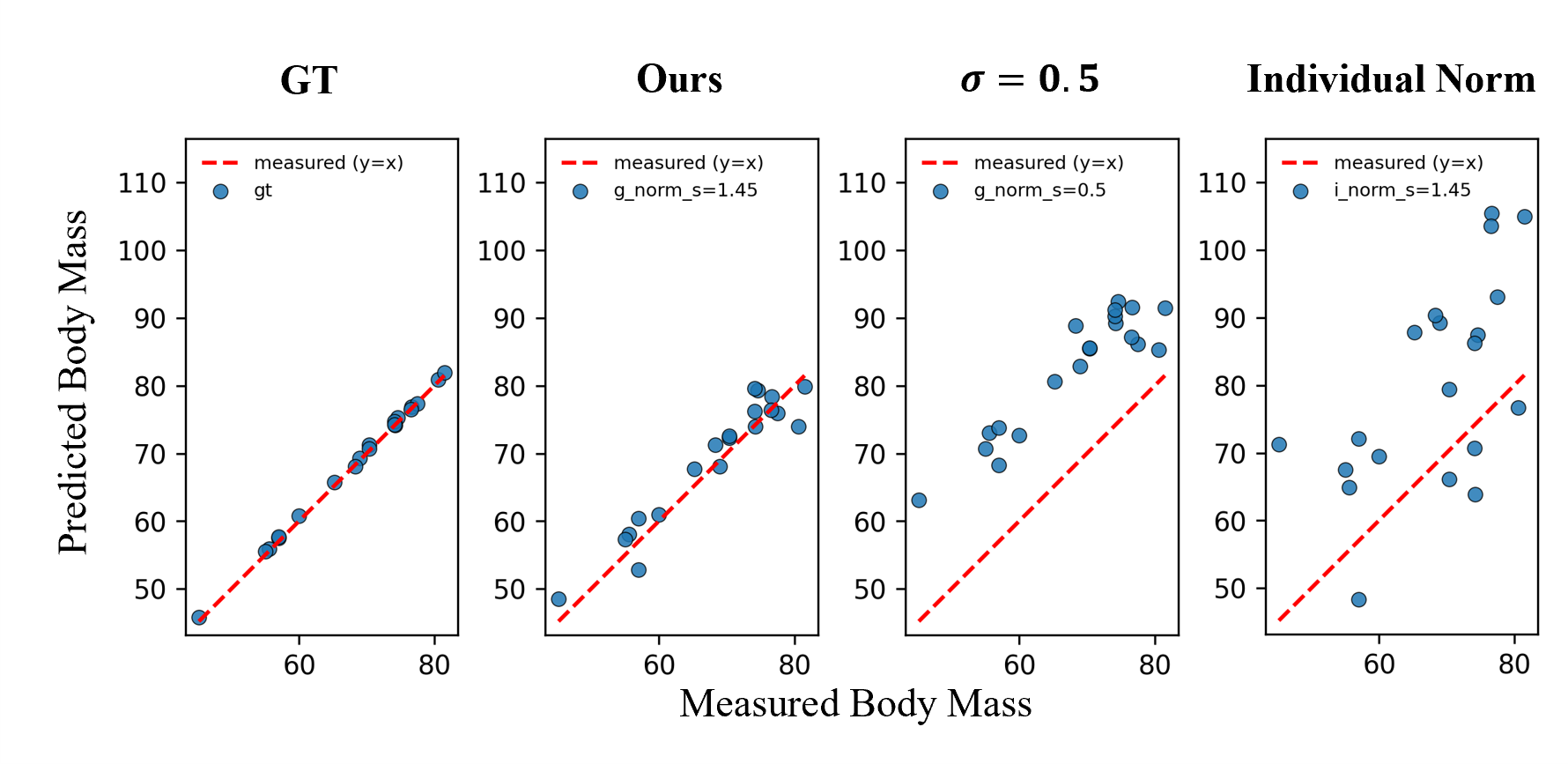}
    \caption{Comparison of \ac{BM} estimation accuracy under different data normalization and pre-processing conditions. The scatter plot correlates predicted \ac{BM} (derived from generated pressure distributions) with actual measurements (red dotted line). Blue points represent predictions from the proposed method trained on $8\,\mathrm{k}$ real samples. Notation: \textit{g\_norm} = global normalization; \textit{i\_norm} = individual normalization; $\sigma$ = Gaussian filter standard deviation (e.g., $\sigma=0.5$ denotes filtering with kernel width 0.5); GT denotes \ac{BM} calculated from ground truth pressure distributions.}
    % \caption{Scatter plot comparing \ac{BM} computed from predicted pressure distributions when proposed method trained with diffent dataset conditions (blue points) with measured \ac{BM} (red dotted identity line); GT denotes \ac{BM} calculated from ground truth pressure distributions. $\sigma=0.5$ indicates pressure distributions in dataset was preprocessed with guassian filter with $\sigma=0.5$ along with global normalization and individual norm indicates each pressure distribution were normalized individually and preprocessed with guassian filter having $\sigma=1.45$; g\_norm = global normalization, i\_norm = individual normalization, and s=0.5 indicates value of $\sigma$. Method wes trained using $8\,\mathrm{k}$ real samples.}
    \label{fig:dataset_conditions}
  \hfill
\end{figure*}

\begin{table*}[h]
\def\arraystretch{1.4}
\centering
\captionsetup{justification=centering}
\caption{Comparison of computational efficiency and model complexity. Metrics are reported for models trained on $8\,\mathrm{k}$ real samples over 100 epochs, with a batch size of 4.}
\label{tab:training_summary}
\begin{tabular}{|l|c|c|c|c|}
\hline
\textbf{Metric} & \citeauthor{ManavarAttFnet} \cite{ManavarAttFnet} & ours & BBDM & LBBDM \\
\hline
Batchsize & \multicolumn{4}{c|}{4} \\ \cline{2-5} 
Model & \textbf{\ac{AttnFnet}-143M} & \textbf{\ac{AttnFnet}-11M} & \textbf{Openai U-Net} & \textbf{Openai U-Net}  \\
Pretrained Network & - & - & - & \textbf{\ac{AttnFnet}-11M} \\
Discriminator & PatchGAN & PatchGAN & - & PatchGAN \\
Input shape & (1, 54, 128) & (1, 54, 128) & (2, 256, 256) & (16, 32, 32) \\
Total Number of Parameters & 143,635,940 & 11,761,018 & 59,303,489 & 258,114,312 \\
Model Parameters Size (MB) & 547.93 & 37.56 & 237.21 & 1032.46 \\
FLOPs & \num{3.92e11} & \num{6.71e9} & \num{1.98e12} & \num{1.29e11} \\ % \cline{1-5}
% FLOPS &  &  &  & \\ % 

training speed (it/s) per GPU & 1.23 & 8.27 & 1.34 & 4.42 \\
Training Speed (sec/epoch) & 838 & 124 & 766 & 232.87 \\
Memory Usage (GB/GPU) & 17.2 & 2.2 & 21.3 & 6.5 \\ % \cline{1-5}

% Power Usage per GPU (W) & 250 & 200 & 242 & 131 \\
% GPU Temperature (\textdegree C) & 80 & 77 & 80 & 60\\

validation speed (it/s) per GPU & 4.99 & 102.34 & 4.1 & 52.50 \\
validation time (sec/2700 img) & 67.58 & 3.27 & 16643 & 1287.02 \\
inference time (sec/img) & 0.11 & 0.024 & 12.63 & 2.76 \\
inference step & 1 & 1 & 200 & 200 \\

\hline
\end{tabular}
\end{table*}

% Model's prediction performance varies as normalization condition varies. In supervised learning, ground truth pressure distribution matter how it is normalized. When normalizing individually pressure distributions leads to weak performance. To overcome that issue, \citeauthor{CleverBodyPressure} \cite{CleverBodyPressure} uses body mass normalization of the pressure distributions of the ground truths. Since, It is difficult to unnormalize generated pressure distributions when not knowing the body mass, they introduced "betanet" to predict body-mass and body-height using latent paramters.

The choice of data normalization strategy significantly impacts the model's predictive performance for pressure distribution. Normalizing each pressure map individually is problematic, as pressure values are co-dependent on external factors like mattress softness. This approach can obscure the relationship between body mass and pressure features, leading to a decline in performance, as illustrated in \cref{fig:dataset_conditions}. While prior work by \citeauthor{CleverBodyPressure}~\cite{CleverBodyPressure} addressed this by normalizing pressure with body mass, their method required an seperate network "Betanet" to predict body mass for denormalization. Betanet only infers \ac{BM} and body-height and does not build any feature body parameter relationship.

Our approach employs a simpler global pressure normalization, where all pressure distributions are scaled into a [0, 1] range based on a single maximum value from the entire dataset. This method proves to be both plausible and suitable for our approach (\cref{fig:dataset_conditions}). However, a challenge arises from pre-processing steps like applying a Gaussian filter. While filtering smooths pressure distributions for better approximation, it can alter the pressure range differently across datasets by reducing low frequency high pressure spike to smoothen the values. For instance, extensive filtering on our real dataset reduces its pressure range to (0,~56)kPa, creating a mismatch with our synthetic data (0,~101)kPa. This discrepancy complicates the generation of valid samples when using \ac{WOL}.

% Current approach uses global pressure normalization where largest pressure value is devided bringing pressure distributions between (0,1). Our approach proves that, without body-mass normalization, generated distributions are plausable and accurate, outperforming baselines.

% Using guassian filter basically smooths the pressure distributions and reduces low freqency high pressure values. However, by doing so we get better approximation as shown in \cref{fig:dataset_conditions} however, synthetic dataset used in this study contains broader pressure range while real dataset used in this study with larger guassian filter reduces pressure range between (0,~56)kPa and this range difference makes it difficult to generate valid samples when using global normalization when using \ac{WOL}. 

% Use of individual normalization is problematic because pressure is also depends upon the softness of the mattress and it leads to mismatch between pressure ranges. When train using individual normalization, model won't be able to build mass - feature realationships and performance reduces (\cref{fig:dataset_conditions}). 

\section{Model Parameters and Training Cost}
\label{sec:model_parameters_and_training_cost}

The computational requirements for our models are summarized in \cref{tab:training_summary}. To improve training speed, we adopted the \ac{AttnFnet} architecture from \citeauthor{ManavarAttFnet}~\cite{ManavarAttFnet} but reduced its size. All models were trained for 100 epochs on a dataset of $8235$ images with a batch size of 4. We utilized Distributed Data Parallel (DDP) on a single node equipped with two Tesla V100-PCIE-32GB GPUs.

The proposed method achieves the fastest training and inference times while requiring the lowest total FLOPs. These efficiency gains demonstrate its capability and potential as a real-time monitoring system.

\section{Model Configurations}
\label{sec:model_configs}

This section provides the detailed hyperparameter settings and architectural configurations used in our experiments to facilitate reproducibility. \Cref{tab:arch_config} details the architecture of the AttnFNet generator and the openai U-Net denoising models (Pixel-space and Latent-space). \Cref{tab:train_config} lists the training hyperparameters, including optimization settings and diffusion scheduling.

\Cref{fig:lbbdm_strategy} illustrates the proposed LBBDM training and inference pipeline for pressure synthesis. \cref{fig:pretraining} shows the pretraining stage, where the \ac{AttnFnet} backbone is optimized using an image reconstruction loss as described in \cref{sec:training_strategy}, together with architectural modifications such as the bottleneck modules \emph{neck1}, \emph{neck2} to reduce computational complexity, and no skip connection to remove feature influence from encoder to decoder. \cref{fig:lbbdm_training} depicts the latent-space denoising training strategy, and \cref{fig:lbbdm_inference} outlines the inference procedure; fire symbols denote learnable layers and cold symbols indicate frozen layers.

We remove the sigmoid activation from the final layer of \ac{AttnFnet}+\ac{ILS} in all experiments. Sigmoid produces bounded probability distributions that cause overestimation of both pressure values and body model \ac{BM} parameters. Additionally, sigmoid saturation at extreme values compresses high/low pressure values into a narrow range, preventing the model from learning physically plausible pressure distributions.

% Training Hyperparameters Table
\begin{table}[h]
    \centering
    \caption{Training Hyperparameters settings across experiments unless stated otherwise.}
    \label{tab:train_config}
    \vspace{0.1cm}
    \begin{tabular}{lc}
        \toprule
        \textbf{Hyperparameter} & \textbf{Value} \\
        \midrule
        \multicolumn{2}{c}{\textit{Optimization}} \\
        Optimizer & Adam \\
        Learning Rate & $1 \times 10^{-4}$ \\
        Adam $\beta_1$ & 0.9 \\
        Batch Size & 16 \\
        Max Epochs & 200 \\
        % LR Scheduler & ReduceLROnPlateau \\
        % Scheduler Factor & 0.5 \\
        % Scheduler Patience & 3000 (steps) \\
        % Min Learning Rate & $5 \times 10^{-7}$ \\
        \midrule
        \multicolumn{2}{c}{\textit{Diffusion Schedule}} \\
        Diffusion Time Steps ($T$) & 1000 \\
        Sampling Steps & 200 \\
        \bottomrule
    \end{tabular}
\end{table}

% Architecture Configuration Table
\begin{table}[h]
    \centering
    \caption{Network Architecture Configurations for the AttnFNet, as well as the U-Net architectures used for the Brownian Bridge Diffusion Model (BBDM) and Latent BBDM (LBBDM) variants.}
    \label{tab:arch_config}
    % \vspace{0.1cm}
   % \resizebox{0.95\linewidth}{!}{%
    \begin{tabularx}{\columnwidth}{lY}
        \toprule
        \textbf{Parameter} & \textbf{Value} \\
        \midrule
        \multicolumn{2}{c}{\ac{AttnFnet}} \\
        \midrule
        Input Resolution & $54 \times 128$  \\
        Image size & $256 \times 256$ \\
        Embedding Dim ($C$) & 264 \\
        Encoding Depth & 8 \\
        Attention Heads & 12 \\
        Patch Size & $16 \times 16$ \\
        Layers with global attention & (1, 2, 5, 8)  \\
        Encoder-decoder skip indices & (4, 6, 8) \\

        \midrule
        \multicolumn{2}{c}{\textit{Diffusion Openai U-Net (BBDM / LBBDM)}} \\
        \midrule
        Input Resolution & $256 \times 256$ / $32 \times 32$ \\
        Base channel width & 64 (Pixel) / 128 (Latent)  \\
        Channel Multipliers & (1, 4, 8) \\
        Attention Resolutions & (32, 16, 8) / (32, 16, 4) \\
        Num Res Blocks & 2 \\
        Attention Heads & 64  \\
        Conditioning & concate  \\
        \midrule
        \multicolumn{2}{c}{\textit{Discriminator}} \\
        \midrule
        Cond/ Uncond discriminator & PatchGAN   \\
        Features per layer & (64, 128, 256, 512) \\
        patch size & $62 \times 62 $ \\
        \bottomrule
    \end{tabularx}
    %}
\end{table}

\section{Realtime Monitoring}
\label{sec:realtime_monitoring}

\begin{figure*}[h]
  \centering
    \includegraphics[width=0.8\linewidth]{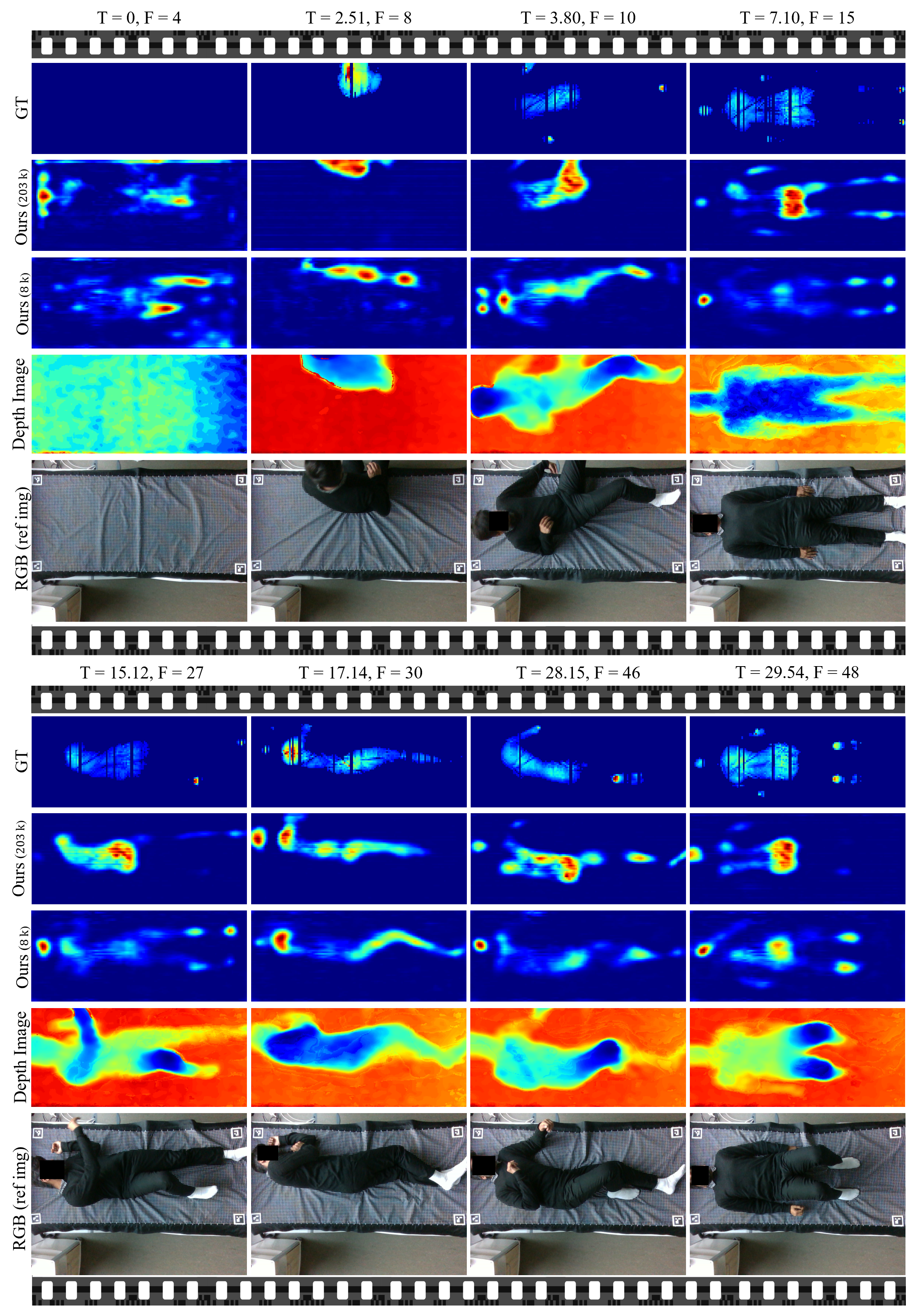}
    \caption{Qualitative evaluation of real-time pressure prediction. We compare the ground truth (GT) pressure distribution against predictions from models trained on $203\,\mathrm{k}$ (Ours $203\,\mathrm{k}$) versus $8\,\mathrm{k}$ (Ours $8\,\mathrm{k}$) samples. Corresponding depth inputs and RGB reference images are shown for context. Timestamps (T) and frame numbers (F) indicate the temporal progression of the trial.}
    \label{fig:realtime_demo}
  \hfill
\end{figure*}

To preliminarily evaluate real-time capability, we conducted a temporal analysis comparing two versions of our method: one trained on $8\,\mathrm{k}$ real samples and another on a larger $203\,\mathrm{k}$ mixed dataset. A qualitative visual comparison is presented in \cref{fig:realtime_demo}.

Our findings show evidence of no overfitting and highlight the need for a diverse training dataset. As shown in \cref{fig:realtime_demo}, the model trained on $203\,\mathrm{k}$ samples demonstrates superior robustness. When the bed is empty, it generates low-amplitude noise. In contrast, the model trained on only $8\,\mathrm{k}$ samples is prone to hallucination, inferring human-like pressure patterns where none exist.

We identified several limitations in the current real-time partial evaluation. First, while \cref{fig:realtime_demo} demonstrates real-time capabilities, the current API (Application Programming Interface) version is not optimized for speed. The current method does not explicitly enforce temporal consistency, which could lead to fluctuations between consecutive frames. A thorough zero-shot quantitative evaluation on unseen environments is needed. Addressing these aspects constitutes a clear direction for future work.

\section{LBBDM Synthetic Pre-Training Results}
\label{lbbdm_synthetic_pretraining}
\begin{figure*}
  \centering
  \begin{subfigure}{\linewidth}
  \centering
    \includegraphics[width=0.98\linewidth]{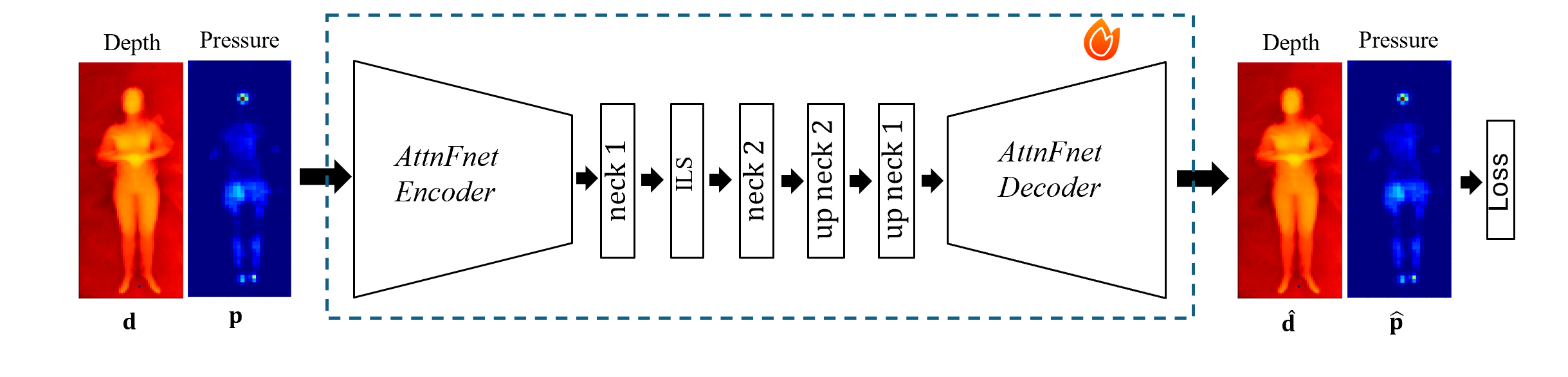}
    %\fbox{\rule{0pt}{2in} \rule{.9\linewidth}{0pt}}
    \caption{pretraining of the \ac{AttnFnet} model with image reconstruction objective.}
    \label{fig:pretraining}
  \end{subfigure}
  \hfill
  \begin{subfigure}{\linewidth}
    \centering
    \includegraphics[width=0.78\linewidth]{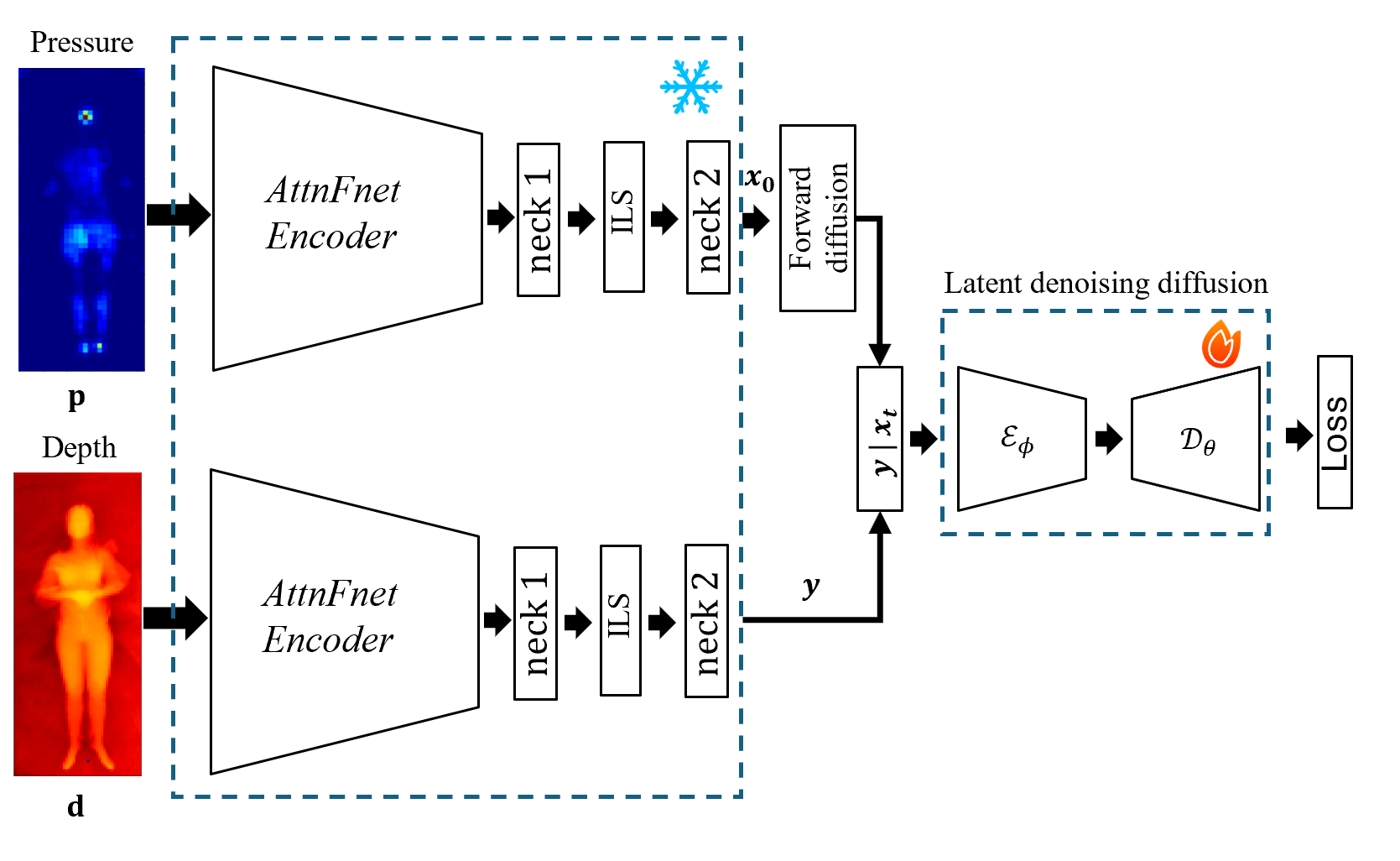}
    %\fbox{\rule{0pt}{2in} \rule{.9\linewidth}{0pt}}
    \caption{training strategy of the denoising model in latent space}
    \label{fig:lbbdm_training}
  \end{subfigure}
  \hfill
  \begin{subfigure}{\linewidth}
    \centering
    \includegraphics[width=0.95\linewidth]{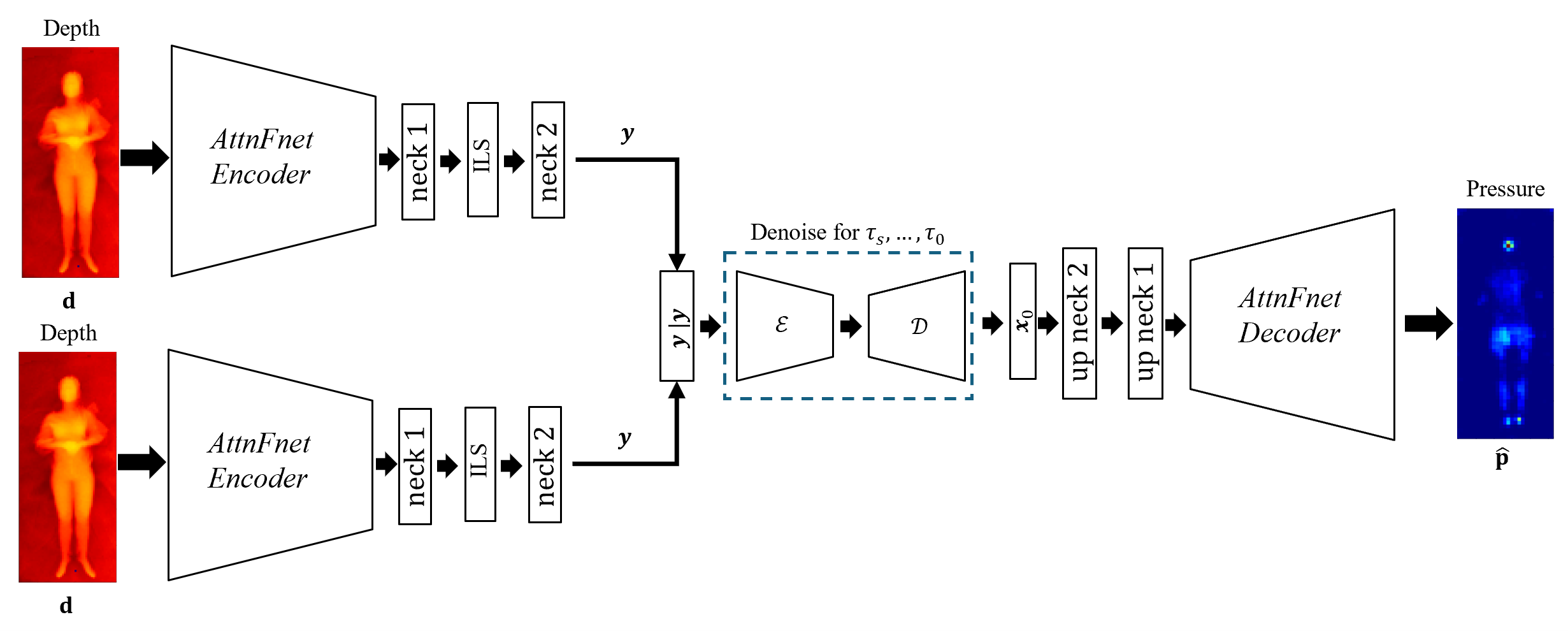}
    %\fbox{\rule{0pt}{2in} \rule{.9\linewidth}{0pt}}
    \caption{schematic diagram of the inference.}
    \label{fig:lbbdm_inference}
  \end{subfigure}
  \caption{schematic of proposed LBBDM training strategy and inference.}
  \label{fig:lbbdm_strategy}
\end{figure*}

This section presents results for a special case of \ac{LBBDM} pre-training. \Cref{fig:lbbdm_strategy} illustrates the training strategy of the conditional \ac{LBBDM} model as a schematic diagram. In contrast to the method described in \cref{sec:training_strategy}, we tested the capability of conditional \ac{LBBDM} to generate real samples when the \ac{AttnFnet} model with \ac{ILS} was pre-trained on synthetic data, while post-training and testing were performed exclusively on real samples.

Pre-training with synthetic data leads to an increase in \ac{MSE}, as shown in \cref{tab:MSE_LBBDM}. These results indicate the importance of real samples during pre-training. A visual comparison of both cases is presented in \cref{fig:visulization_sup}.

 % We attribute the higher \ac{MSE} to a mismatch in the pressure range between synthetic and real data: samples predicted by the \ac{AttnFnet} decoder tend to produce idealized pressure distributions, whereas real measurements contain non-ideal patterns and pressures up to 56~kPa (see \cref{sec:eff_data_norm}), which leads to larger errors.

\begin{table}[h]
  \centering
  \begin{tabularx}{\columnwidth}{lYXYYX}
    \toprule 
    Method & Ucov & cov ($1$mm) & cov ($3$mm) & overall\\
    \midrule
    LBBDM $8\,\mathrm{k}$ & 0.9047 & 1.0590 & 1.0803 & 1.0149 \\
    LBBDM-synth & 1.0463 & 1.2318 & 1.2514 & 1.1765 \\
    \bottomrule
  \end{tabularx}
  \caption{Test set \ac{MSE} in $kPa^2$ for uncovered (Ucov), covered at $1$ mm, covered at $3$ mm, and overall poses. The results compare \ac{LBBDM} pre-trained on real samples (\ac{LBBDM} $8\,\mathrm{k}$) versus synthetic-only pre-training (\ac{LBBDM}-synth).}
  \label{tab:MSE_LBBDM}
\end{table}

\section{Additional Results}
\label{sec:additional_results}

\begin{figure*}[h]
  \centering
    \includegraphics[width=0.95\linewidth]{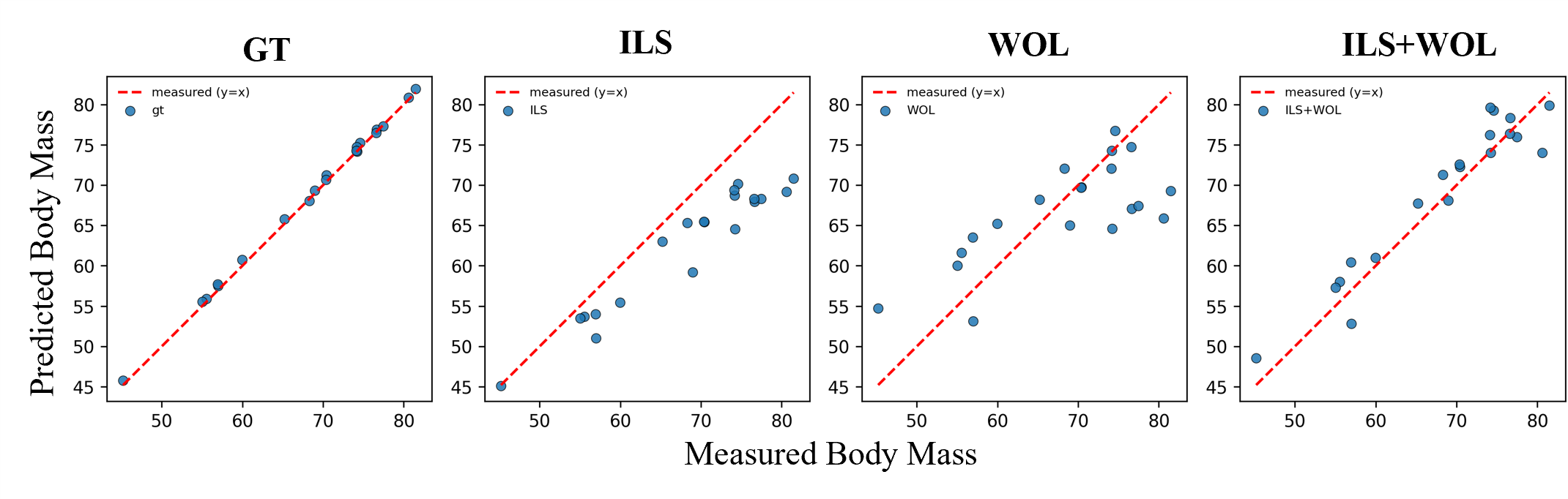}
    \caption{Impact of proposed components on \ac{BM} estimation consistency. The scatter plot compares predicted \ac{BM} ($\hat{\text{BM}}$) against measured \ac{BM} (red dotted line) and ground truth (GT) values (blue dots align with red dotted line) under different training configurations (blue points). \textbf{ILS}: Informed Latent Space only; \textbf{WOL}: Weight Optimization Loss only; \textbf{ILS+WOL}: Combined approach. All models were trained on $8\,\mathrm{k}$ real samples.}
    \label{fig:component_impact}
  \hfill
\end{figure*}

\Cref{fig:component_impact} illustrates the individual and combined impact of \ac{ILS} and \ac{WOL} on \ac{BM} plausibility. The use of \ac{ILS} enhances \ac{BM} consistency, whereas \ac{WOL} ensures the generated distribution aligns closely with the participant's ideal \ac{BM}. Integrating both components results in improved consistency and significantly reduced \ac{BM} estimation error.

\begin{figure*}[h]
  \centering
    \includegraphics[width=0.75\linewidth]{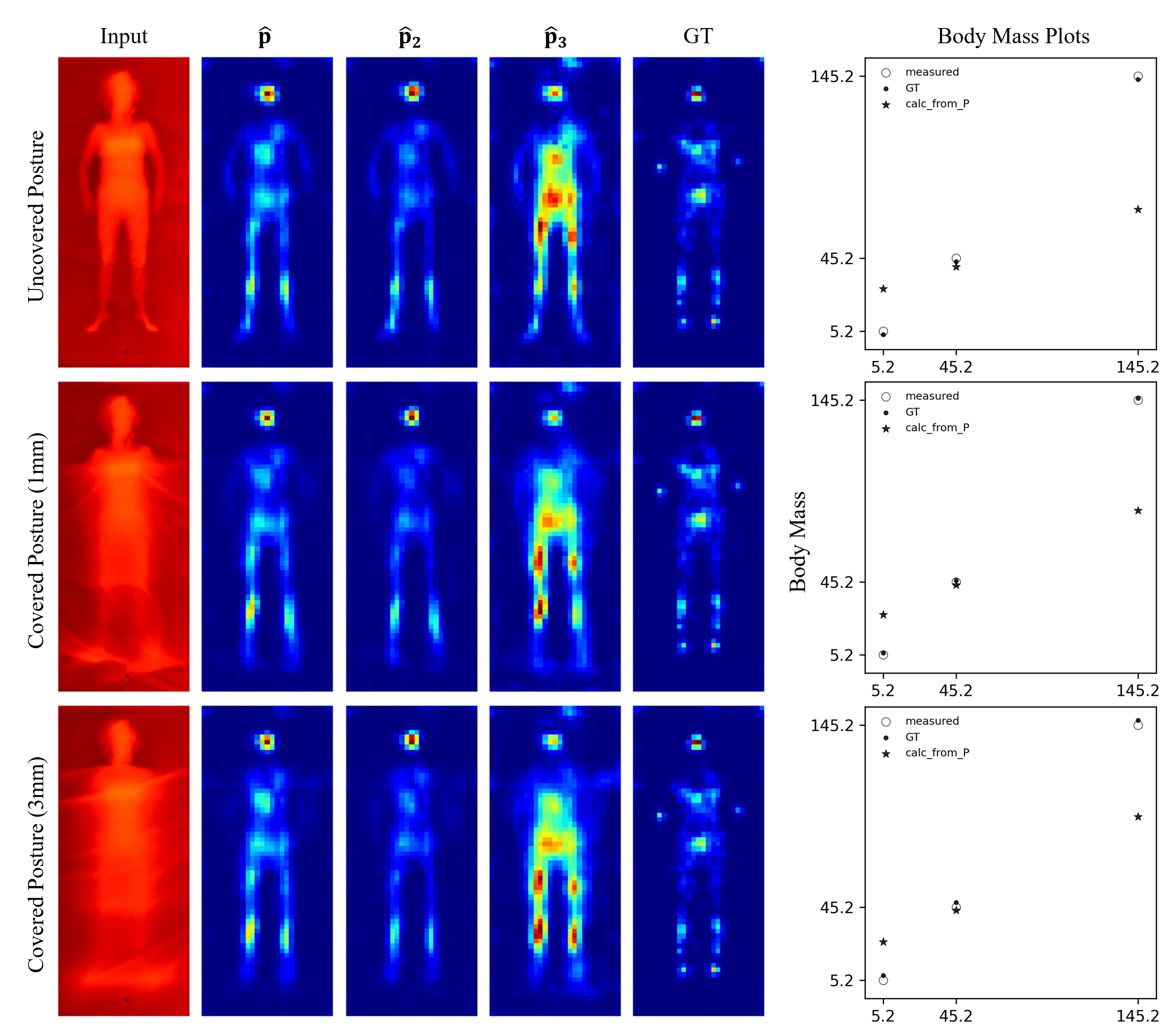}
    \caption{Visualization of diverse pressure distributions generated from a single depth input using different anthropometric prompts. \textbf{Left:} Predicted pressure maps $\mathbf{\hat{p}}$, $\mathbf{\hat{p}_2}$, and $\mathbf{\hat{p}_3}$ correspond to varying mass inputs ($m=45.2$, $5.2$, and $145.2$ kg, respectively) while holding height ($h=1.51$ m) and gender ($g=\text{Female}$) constant. \textbf{Right:} Corresponding computed Body Mass (\ac{BM}) plots. Models were trained on $203\,\mathrm{k}$ mixed (real + synthetic) samples.}
    \label{fig:multi_pred}
  \hfill
\end{figure*}

\Cref{fig:multi_pred} demonstrates \ac{BM} adaption capability of \ac{ILS} via anthropometric conditioning. Using a fixed depth input, the model synthesized three distinct pressure distributions: $\mathbf{\hat{p}}$ using actual parameters ($m=45.2,\ h=1.51,\ g=\text{Female}$), $\mathbf{\hat{p}_2}$ with reduced mass ($m=5.2,\ h=1.51,\ g=\text{Female}$), and $\mathbf{\hat{p}_3}$ with increased mass ($m=145.2,\ h=1.51,\ g=\text{Female}$). Reducing the input mass decreases the overall inferred pressure, while increasing it amplifies pressure, particularly in the pelvic region where high loads are expected. This adaptability suggests the system can accommodate patient weight fluctuations in real-world deployments by adjusting conditioning inputs, without other adjustments to the model.

The derived \ac{BM} values are plotted in \cref{fig:multi_pred}. While the \ac{BM} calculated from $\mathbf{\hat{p}_2}$ and $\mathbf{\hat{p}_3}$ follows the trend of the conditioning inputs, it does not perfectly match the target values. This discrepancy arises because the input depth profile remains static, reflecting the subject's original profile—which conflicts with the synthetic mass prompts. Conversely, the model depth - anthropometric relationship remains intact when using actual parameters $\mathbf{\hat{p}}$. Future work can lead to validating this method against ground-truth data from controlled weight gain and loss scenarios.

\begin{figure*}[h]
  \centering
    \includegraphics[width=0.8\linewidth]{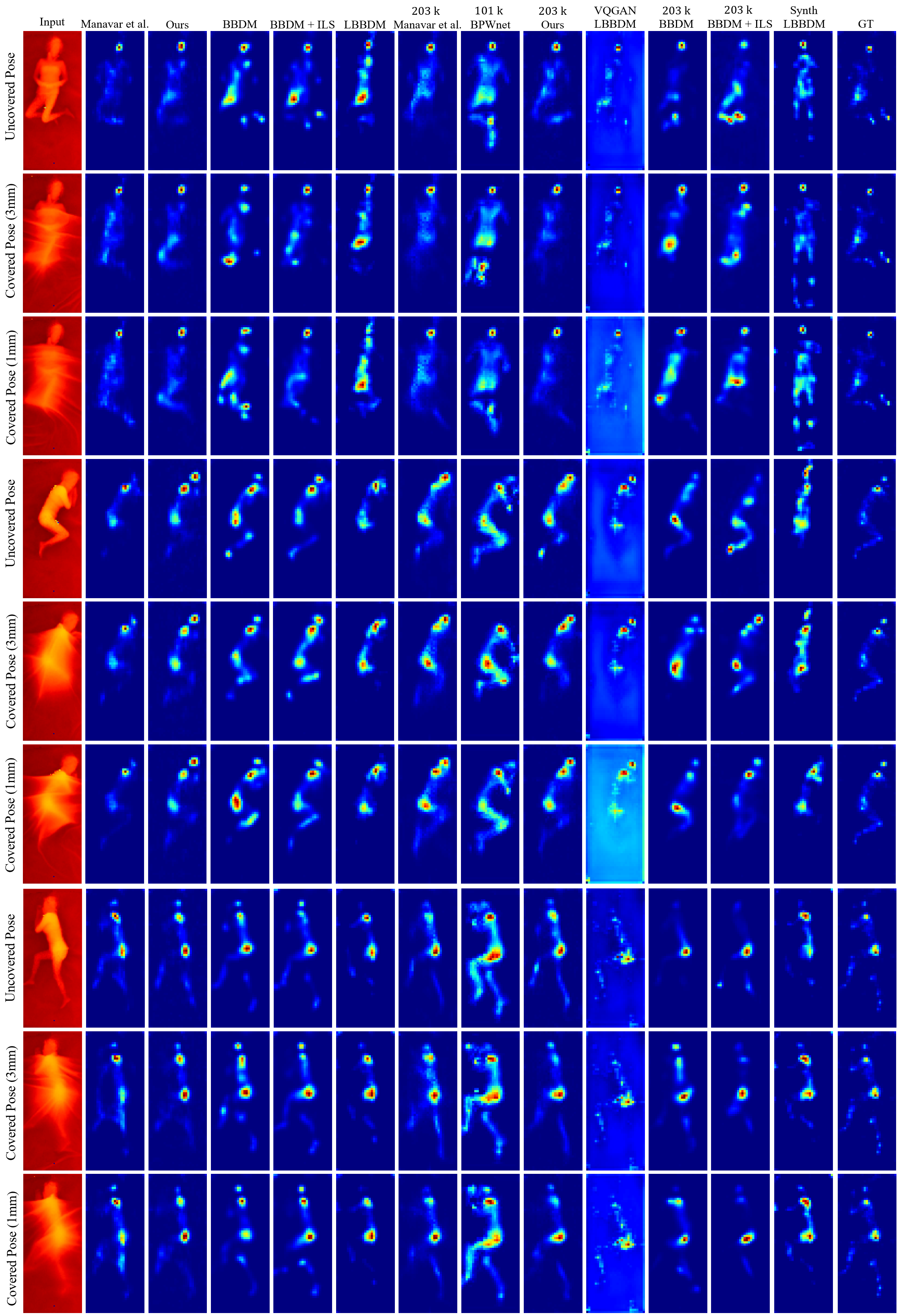}
    \caption{Qualitative comparison of predicted pressure distributions across different methods. Rows represent distinct test samples, while columns display the outputs from competing models against the ground truth (GT). VQGAN represents pretraining happened with the VQGAN model on the CelebHQ dataset, synth represents \ac{AttnFnet}-11M pretrained on synthetic data and post trainining preformed on real data.}
    \label{fig:visulization_sup}
  \hfill
\end{figure*}

Qualitative results are shown in \cref{fig:visulization_sup}, including comparisons with \ac{LBBDM} with a pretrained VQGAN and \ac{LBBDM} with pretraining on synthetic data. Complex scenarios, such as resting poses covered by a blanket, challenge all evaluated models. In these cases, baselines like \ac{BBDM} and BPWnet often hallucinate incorrect postures. In contrast, the proposed method reduces error by blurring unseen regions rather than generating misleading features.

\begin{figure*}[h]
  \centering
    \includegraphics[width=\linewidth]{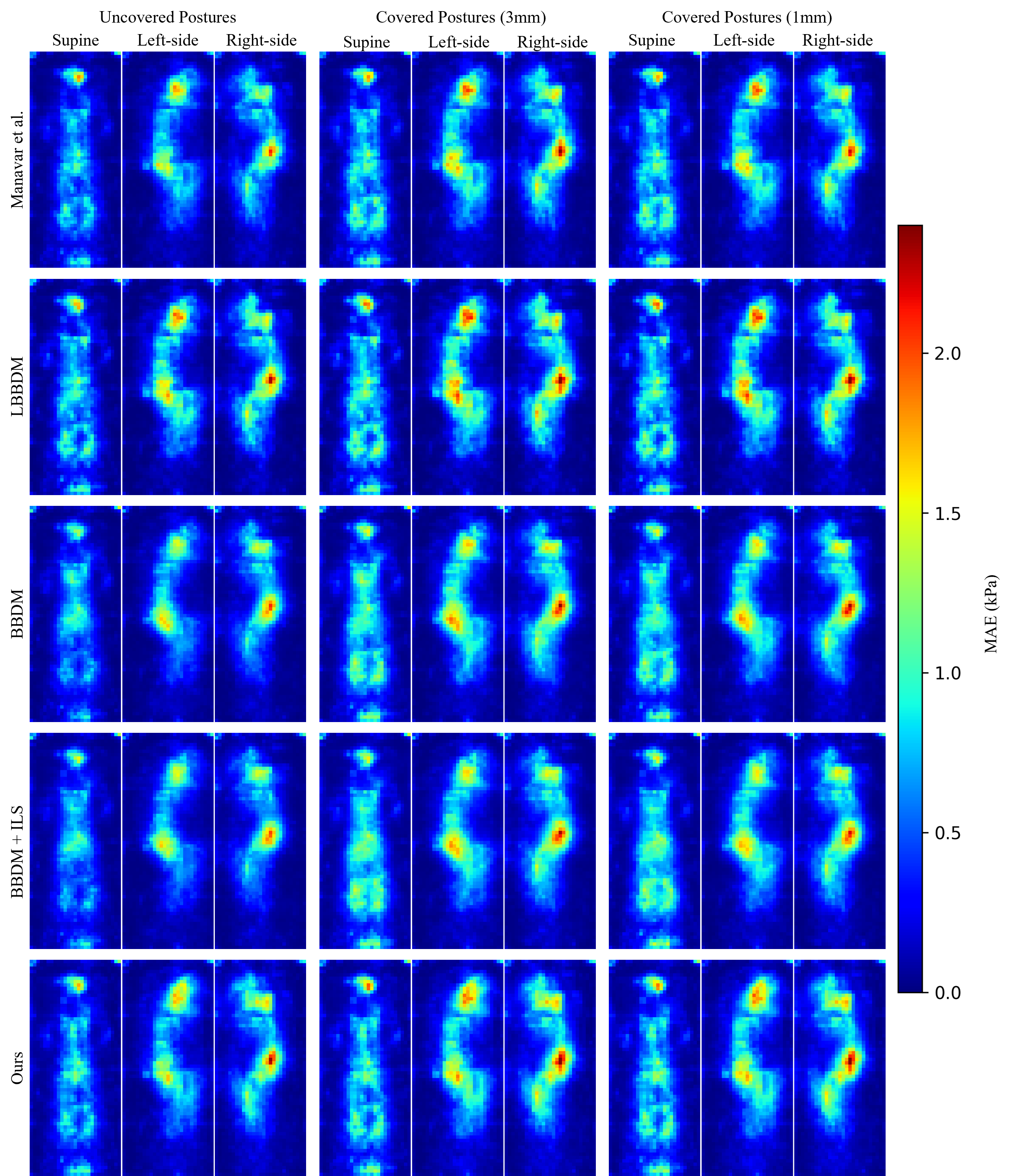}
    \caption{Analysis of \ac{MAE} posture deviations across supine, left-side, and right-side positions. Rows correspond to different training methods, while columns represent varying cover conditions (e.g., no cover vs. blanket). All models were trained on $8\,\mathrm{k}$ real samples.}
    \label{fig:deviations}
  \hfill
\end{figure*}

\Cref{fig:deviations} quantifies the \ac{MAE} across supine, left-side, and right-side postures under three conditions. Consistent with expectations, the pelvic region exhibits the largest errors across all methods. Uncovered postures yield the lowest deviations, while scenarios involving a 3mm blanket result in the highest deviations due to occlusion. The proposed method and \ac{BBDM} + ILS leads to lower \ac{MAE} in the pelvic region.

%% file: sec/7_abbreviations.tex
\begin{acronym}[vitpup]
    \acro{NLP}[\textsc{NLP}]{Natural Language Processing}
    \acro{FCN}[\textsc{FCN}]{Fully Convolutional Network}
    \acro{GAN}[\textsc{GAN}]{Generative Adversarial Network}
    \acro{CNN}[\textsc{CNN}]{Convolutional Neural Network}
    \acro{ViT}[\textsc{ViT}]{Vision Transformer}
    \acro{SAM}[\textsc{SAM}]{Segment Anything Model}
    \acro{SSIM}[\textsc{SSIM}]{Structural Similarity Index}
    \acro{MSE}[\textsc{MSE}]{Mean Squared Error}
    \acro{MAE}[\textsc{MAE}]{Mean Absolute Error}
    \acro{cGAN}[\textsc{cGAN}]{Conditional Generative Adversarial Network}
    \acro{LSGAN}[\textsc{LSGAN}]{Least Square Generative Adversarial Network}
    % \acro{vitpup}[\textsc{vitpup}]{Vision Transformer Progressive Up-Scalling}
    \acro{MLP}[\textsc{MLP}]{Multi-Layer Perceptron}
    \acro{MPPA}[\textsc{MPPA}]{Mean Pixel Prediction Accuracy}
    \acro{MSSIM}[\textsc{MSSIM}]{Mean Structural Similarity Index}
    \acro{MFID}[\textsc{MFID}]{Mean Fréchet Inception Distance}
    \acro{MPSNR}[\textsc{MPSNR}]{Mean Peak-Peak Signal-to-Noise Ratio}
    \acro{PPA}[\textsc{PPA}]{Per Pixel Accuracy}
    \acro{FID}[\textsc{FID}]{Fréchet inception distance}
    \acro{PSNR}[\textsc{PSNR}]{Peak-Peak Signal-to-Noise Ratio}
    \acro{IQR}[\textsc{IQR}]{interquartile range}
    \acro{OFDI}[\textsc{OFDI}]{Occlusion Free Depth Images}
    \acro{PPress}[\textsc{PPress}]{Pre-processed Pressure distribution}
    \acro{AttnFnet}[\textsc{AttnFnet}]{Attention Feature Network}
    \acro{IOU}[\textsc{IOU}]{Intersection Over Union}
    \acro{MIOU}[\textsc{MIOU}]{Mean Intersection Over Union}
    \acro{SLP}[\textsc{SLP}]{Systematic Lying Postures}
    \acro{PEye}[\textsc{PEye}]{Pressure Eye}
    \acro{ILS}[\textsc{ILS}]{Informed Latent Space}
    \acro{WOL}[\textsc{WOL}]{Weight Optimization Loss}
    \acro{BBDM}[\textsc{BBDM}]{Brownian Bridge Diffusion Model}
    \acro{LBBDM}[\textsc{LBBDM}]{Latent Brownian Bridge Diffusion Model}
    \acro{BM}[\textsc{BM}]{Body-Mass}
\end{acronym}